# Guided acoustic wave sensors for liquid environments

## C. Caliendo[*1], M. Hamidullah[1,2]

[1]Institute for Photonics and Nanotechnologies, IFN-CNR, Via Cineto Romano 42, 00156 Rome, Italy

[2]Department of Information Engineering, Electronics and Telecommunications, University of Rome La Sapienza, 00156 Rome, Italy

*Corresponding author; email: cinzia.caliendo@cnr.it

**ABSTRACT.** Surface acoustic wave (SAW) based sensors for applications to gaseous environments have been widely investigated since the last 1970s. More recently, the SAW-based sensors focus has shifted towards liquid-phase sensing applications: the SAW sensor contacts directly the solution to be tested and can be utilized for characterizing physical and chemical properties of liquids, as well as for biochemical sensor applications. The design of liquid phase sensors requires the selection of several parameters, such as the acoustic wave polarizations (i.e., elliptical, longitudinal and shear horizontal), the wave-guiding medium composition (i.e., homogeneous or non-homogeneous half-spaces, finite thickness plates or composite suspended membranes), the substrate material type and its crystallographic orientation. The paper provides an overview of different types of SAW sensors suitable for application to liquid environments, and intents to direct the attention of the designers to combinations of materials, waves nature and electrode structures that affect the sensor performances.

## 1. Introduction

The basic structure of a Surface Acoustic Wave (SAW) sensor includes a piezoelectric substrate and a pair of interdigitated transducers (IDTs) photolithographically patterned onto the free surface of the piezoelectric wafer [1]. The piezoelectric substrate can be a *bulk* single crystal substrate (such as quartz, $LiNbO_3$ or $LiTaO_3$, to name just a few), with thickness h larger than of the acoustic wavelength $\lambda$ (h >> $\lambda$), or a *thin film* (such as ZnO or AlN) with thickness h < $\lambda$, grown onto a non-piezoelectric *bulk* substrate (such as silicon, sapphire, glass, or diamond, to cite just a few). In SAW devices, the acoustic wave travels at the surface of the propagating medium and its energy is confined within one wavelength of depth; it follows that the SAW properties (wave velocity and amplitude) are highly affected by any physical and chemical changes that occur at the surface of the propagating medium and/or at the adjacent medium, when the SAW device interacts with an external environmental stimuli (such as humidity, temperature, and pressure, to cite just a few). In the presence of a liquid environment, the wave properties can be perturbed by the changes of the electrical (conductivity and permittivity) and mechanical (mass density and viscosity) properties of the liquid contacting the sensor surface, or by the anchorage of a mass onto the sensor surface.

The environment can interact with the SAW directly or by means of a thin *sensing* membrane that, for some specific applications, covers the wave propagation path between the two IDTs and is in direct contact with the environment to be tested. If the membrane is an insulating material, it can cover the entire SAW device surface, including the IDTs, while if it is conductive, it is positioned in between the two IDTs. As a consequence, part of the SAW energy is distributed into the sensing membrane and any change in its physical properties affects either or both the wave velocity and propagation loss, giving rise to a detectable output signal (a frequency and/or insertion loss shift) that represents the sensor response. The affinity of the membrane towards a specific target analyte is a fundamental prerequisite as it can drive the sensor selectivity towards a specific application. In gas sensing applications, the sensing membrane can be a thin Pd film [2, 3], a thin lead phthalocyanine (PbPc) film [4] a graphene-like nano-sheet [5], a calixarene layer [6] or a polyethynyl-fluorenol layer [7], to cite just few examples, to detect $H_2$, $NO_2$, carbon monoxide, organic vapors or simply the relative humidity of the surrounding environment. In liquid phase sensing applications, the membrane can be poly(isobutylene) (PIB), poly(epichlorohydrin) (PECH), or poly(ethyl acrylate) (PEA) to test toluene, xylenes, and ethyl benzene solutions [8], or polysiloxane film containing acidic functional groups for detection of organic amines in aqueous phase [9], or macrocyclic calixarenes for the detection of organic pollutants in drinking water [10]. The SAW sensors described in [11] were fabricated and derivatized with a rabbit polyclonal IgG antibody, which selectively binds to E. coli O157:H7. A dual channel SAW biosensor for the simultaneous detection of Legionella and Escherichia coli was fabricated using a novel protocol of coating bacteria on the sensor surface prior to addition of the antibody [12]. Reference [13] shows an overview of 20 years-worldwide developments in the field of SAW-based biosensors for the detection of biorelevant molecules in liquid media.

Whatever the sensing membrane is, it is important to underline that the design of the device can significantly affect the performances of the sensor since its sensitivity is also dependant on their wave-type and polarization, on the electroacoustic coupling configuration, on the materials thickness and crystallographic orientation.

SAW devices are manufactured with semiconductor integrated circuits technologies that include the metal and piezoelectric layers deposition techniques (such as sputtering, pulsed laser deposition or thermal evaporation), optical or electronic lithography, and lift-off process to pattern the IDTs and the sensing membrane. The selection of the substrate (the material type, crystallographic cut and thickness), the IDTs design (metal fingers width and spacing, number of fingers pairs, aperture, IDTs centre-to-centre spacing, to cite just a few), as well as the selection of the electroacoustic coupling configuration affect the characteristics of the electroacoustic devices (such as the operating frequency, quality factor, insertion loss) in such a way as to enhance the sensitivity toward the environmental parameters changes, regardless of the type of the adopted sensing membrane.

The SAW sensors are implemented on piezoelectric substrates showing high electroacoustic coupling, such as ZnO, $LiNbO_3$ and $LiTaO_3$, that ensure a high sensor sensitivity, sometimes at the price of a moderate temperature stability. It can happen that the thermal gradient can lead to frequency shifts which are comparable to the sensor response to the measurand. The selection of temperature stable cuts of the piezoelectric substrates, such as the quartz ST-cut, is one way to improve the temperature-frequency stability of the SAW sensors. A dual sensors configuration, that includes the active sensor and a reference sensor [14-16], can be designed to compensate common mode influences such as humidity or aging, other than temperature, that affect both the active device, coated, for example, with a selectively adsorbing membrane, and the uncoated one that acts as a reference device. Alternatively, a temperature compensated configuration, i.e. a multi-layered structure including opposite-sign temperature coefficients of delay (TCD) materials, with the proper thicknesses [17], can be designed to cancel the device spurious responses due to the thermal drift [18, 19].

The present paper gives a survey of the electroacoustic devices based on the propagation of elastic waves travelling at the plane surface of *half-spaces* and within finite thickness *plates*; the common characteristic of these waves is the ability to travel in a medium contacting a liquid environment, without suffering excessive energy loss. The paper is structured as follows: in Section 2 we consider features of the pseudo surface acoustic waves (PSAWs) and high velocity PSAWs (HVPSAWs), showing in-plane polarization and travelling at the surface of piezoelectric substrates; results of numerical calculations and finite element modelling of the PSAW and HVPSAW-based sensors for liquid environments are presented. In Section 3 the features of Love wave-based sensors are studied for different combinations of substrate and guiding layer materials, with the varying layer thickness. In Section 4 the features of shear horizontal acoustic plate modes (SHAPMs)-based sensors are studied for different plate thicknesses, material types and modes order. In Section 5 Lamb wave-based sensors are studied, including the quasi-longitudinally polarized fundamental and higher order modes, the fundamental symmetric mode and the quasi Scholte wave. Section 6 concludes our paper.

## 2. Pseudo SAW and high velocity PSAW sensors

Surface acoustic waves (SAWs) are elastic waves that travel at the free surface of a half-space and are confined within one acoustic wavelength λ in depth. The physical motion of the SAW is mechanically associated with an elliptical displacement of the surface that is characterized by one out-of-plane particle displacement component, $U_3$, and two in-plane components, $U_2$ and $U_1$, normal and parallel to the wave vector $k = \frac{2\pi}{\lambda}$. When travelling in a piezoelectric medium, the SAW strain field is accompanied by a travelling electric potential wave: the linear electromechanical coupling effect in piezoelectric materials enables the inter-conversion between electrical and acoustic signals. As a consequence, the excitation and detection of SAWs, as well as of any types of plate waves on piezoelectric substrates, is accomplished by means of metal interdigitated electrodes (IDTs) as first reported by White and Voltmer [20] Figure 2.1 shows the schematic of a single-electrode-type IDT with uniform fingers spacing and constant overlap: several metal strips are aligned and connected to the bus bars with a periodicity corresponding to the λ. The fingers width and spacing are equal to λ/4; the total length of the IDT is L = (N − ¼)·λ, being N the number of finger pairs; W is the electrodes overlapping (the IDT aperture); d =W/λ is the IDT directivity. The frequency of the propagating wave is $f = v/\lambda$, where $v$ is the velocity of sound in the half-space material. The SAWs propagate in both directions away from the IDT, along the propagation axis: thus, the inherent loss of the two IDTs is equal to 3 dB.

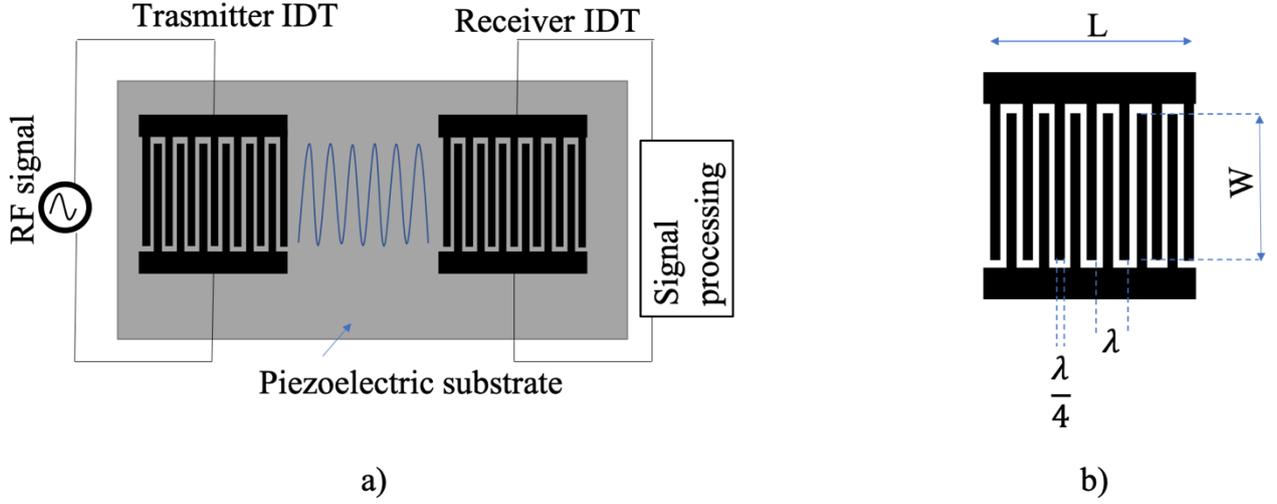

Figure 2.1: a) schematic view of the fundamental structure of a SAW device including a launching and a receiving transducer; b) interdigital transducer.

When a RF voltage is applied between the two bus bars of the transmitting IDT, a periodic strain is generated. As a travelling electric field is associated to the SAW, the metal strips of the receiving IDT will detect the SAW-induced charges. By changing the fingers overlapping, the IDT aperture, N and the spacing of the metallic fingers, the SAW bandwidth and the directivity can be changed. A complete description of the theory and modelling of IDTs employed for SAW excitation and detection can be found in Reference [21]. The electromechanical coupling factor $K^2$ is a measure of the electric-to-acoustic energy conversion efficiency; it is represented by the fractional change in wave velocity due to surface metallization, $K^2 = 2 \cdot \left[ \dfrac{(v_{free} - v_{met})}{v_{free}} \right]$, being $v_{free}$ the free surface wave velocity and $v_{met}$ the velocity on the metallized surface. The $K^2$ values of the SAWs travelling along common piezoelectric substrates are equal to 0.16, 0.75, 4.8 and 5.31% for ST-quartz, 112°x-y LiTaO$_3$, zx- and 128°y-x LiNbO$_3$, respectively [22]. Due to the anisotropic properties of the piezoelectric materials, the $K^2$ depends on the substrate crystallographic cut and wave propagation direction. The electrical impedance of the IDT depends on several factors such as $K^2$, the dielectric permittivity of the substrate, and the geometry of the IDT; it must match as closely as possible that of external components (50 Ω) and be resistive. IDT apertures of less than approximately 30·λ are inadvisable, as the transducer can diffract the acoustic beam resulting in an acoustic beam considerably diverging before reaching the output IDT [23]. The metal used to fabricate the IDTs is generally a low-resistance material highly adhesive to the substrate surface, with thickness around 0.1 μm, in order to make the transduction process more efficient; Al, Au, Cu, and Mo are few examples of commonly used metals: some of these materials require the presence of a thin adhesive inter-layer (Ti or Cr) to improve the adhesion to the substrate or to avoid the diffusion of the metal into the substrate. The cost is also an important factor that drives the choice of the IDT metal type: Cr/Al appears to be the best choice with a good balance of relatively low resistivity, low cost, and good surface adhesion. In harsh environment applications (such as high temperature or corrosive environments), the metal used to fabricate the IDTs should exhibit a high electrical conductivity, high melting temperature, a good resistance to oxidation and chemical inertness. Iridium, Rhodium and Platinum are few examples of metals suitable for harsh environment applications [24-27].

In contrast to SAWs, which are polarized perpendicularly to the surface, the pseudo SAWs (PSAWs) and high velocity PSAWs (HVPSAWs) are predominantly in-plane polarized, that makes them suitable for low loss propagation under a liquid environment. The three SAW displacement components decay exponentially with the depth, while the PSAWs and HVPSAWs have both decaying and radiating components; the latter component radiates power into the half-space, thus resulting in an attenuation of the field amplitudes as the wave propagates. If the contribution from the radiating terms is sufficiently small, these two pseudo waves are observed as in standard SAW devices. The PSAW usually has the $U_2$ component as the dominant one ($U_2 \gg U_1$, $U_3$ at the half-space surface), while the HVPSAW usually has the longitudinal component $U_1$ as the dominant term ($U_1 \gg U_2$, $U_3$ at the half-space surface) [28]. For specific crystallographic cuts and wave propagation directions in the most common piezoelectric substrates, piezoelectrically-active PSAWs and HVPSAWs travel with minimum propagation loss and, since they are both in-plane polarized, they are suitable to work in contact with liquid. Three dimensional (3D) eigenfrequency FEM analysis was performed using COMSOL Multiphysics® Version 5.2 to explore the field shape of the SAW, PSAW and HVPSAW travelling at the surface of a ST-x quartz substrate (Euler angles 0° 132.75° 0°). Figure 2.2 shows the 3D primitive SAW cell as considered in the analysis: beneath the single wavelength cell is a perfectly matched layer (PML) at the bottom for capturing losses related to bulk wave radiation, in order to simulate the half-space. The primitive SAW cell has two periodic and two continuity boundary conditions applied on the sidewalls. The Al electrodes are 0.1 µm thick, with pitch of p = 5 µm ($\lambda$ = 20 µm). The IDT fingers width-to-spacing ratio was set to 1. The base material is ST-x quartz.

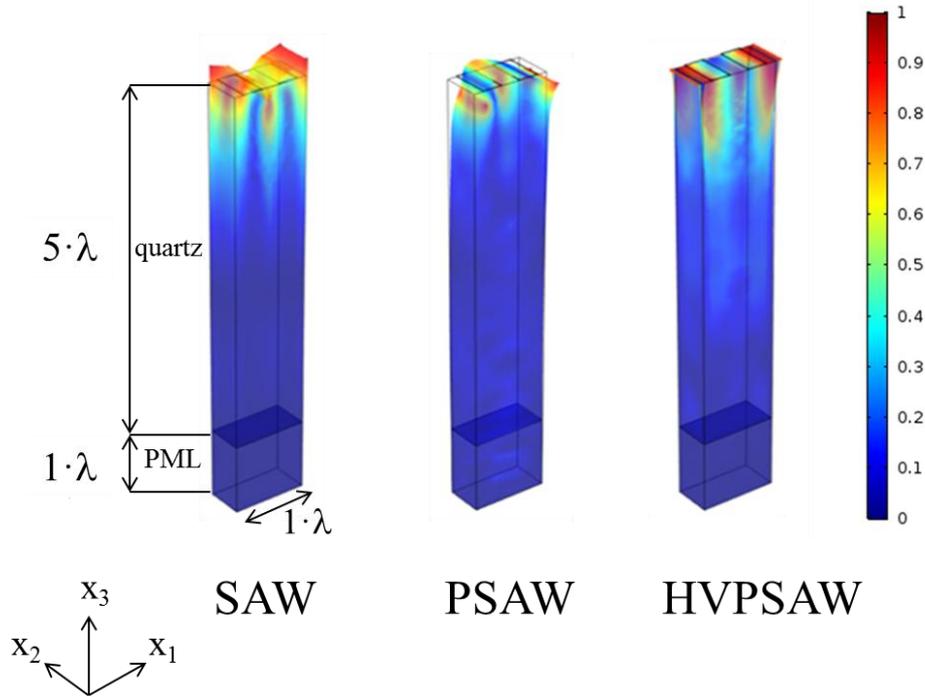

Figure 2.2: The 3D field profile of the SAW, PSAW and HVPSAW travelling along the ST-x quartz substrate in air. The colour density bar is representative of the relative particle displacement: the dark blue represents zero displacement and red the maximum displacement.

From figure 2.2 it can be clearly observed that, unlike what happens for the SAWs, the particle motion of the PSAW and HVPSAW is contained in the surface plane of the propagating medium since the shear vertical displacement component is very small for both the two waves.

Attractive properties of PSAWs and HVPSAWs are the high velocity (close to that of the transverse and longitudinal bulk acoustic wave, respectively), low propagation loss and high electromechanical coupling coefficient [28].

The sensors based on PSAWs and HVPSAWs can measure the mechanical (mass density and viscosity) and electrical (conductivity and relative permittivity) property changes of the liquid that contacts the wave path directly, without covering the sensor surface with any selective film. The sensor surface is in direct contact with the liquid test bath whose thickness greatly exceeds the penetration depth of the wave mode excited by the IDTs. The sensor response is caused by the mechanical and electrical boundary conditions changes resulting from the perturbations the adjacent medium undergoes. If the wave propagation path is metallized and electrically shorted, only the liquid mechanical properties will affect the sensor response, as only the particle displacement component interacts with the adjacent liquid. If the *bare* acoustic path is in direct contact with the liquid, the sensor will be sensitive also to the electrical properties of the liquid as both the wave electrostatic potential and particle displacement interact with the liquid, and two perturbations occur. The electrical perturbation can be discriminated by detecting differential signals between two delay lines.

The absolute value of the normalized admittance Y vs frequency curves for the three modes propagating in ST-x quartz, in air and in water, were calculated by a frequency domain study, and the curves are shown in Figure 2.3. Three peaks are clearly visible when the surface contacts the air (black curve): they correspond to the SAW, PSAW and HVPSAW whose velocities $v = f \cdot \lambda$ are equal to 3167, 5083 and 5751 m/s, respectively. The PSAW has a propagation loss higher than that of the SAW and HVPSAW [29]. The red curve of figure 2.3 corresponds to ST-quartz half-space contacting the water. The SAW, which has a large displacement component normal to the substrate, is almost totally damped by the water as expected; the PSAW shows nearly no damping while the HVPSAW is affected by a small attenuation. As the vertical displacement component $U_3^{surf}\big/U_1^{surf}$ of the HVPSAW, normalized to the $U_1$ at the surface, is about four times that of the PSAW, the former wave is more dampened by the water than the latter.

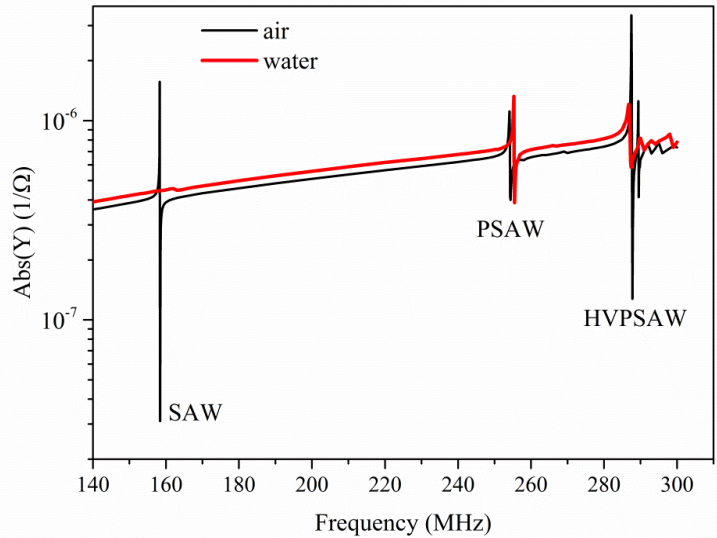

Figure 2.3: The absolute value of the normalized admittance Y vs frequency curves of the SAW, PSAW and HVPSAW travelling at the surface of the ST-x quartz substrate in air and in water.

A two dimensional (2D) COMSOL simulation of the HVPSAW displacement profile inside the ST-x quartz is plotted in figure 2.4a: the blue and the green curves represent the $U_1$ and $U_3$ HVPSAW displacement components; the abscissa is the normalized depth (for $\lambda = 20$ μm). Figure 2.4b shows the 2D representation of the HVPSAW total displacement. The substrate and liquid depths (120 μm) are equal: the ST-x quartz extends from -120 to 0 μm of the abscissa values, while the liquid half-space extends from 0 to 120 μm abscissa value. The liquid was modelled as a linear isotropic viscoelastic material with independent elastic constants; the bulk modulus and the dynamic viscosity were extracted from Reference [30]. When the quartz is contacted by the water, the surface-normal displacement component of the HVPSAW (the blue line) generates compressional waves in the liquid phase: the power dissipated leads just to a small attenuation since the wave is predominantly in-plane polarized.

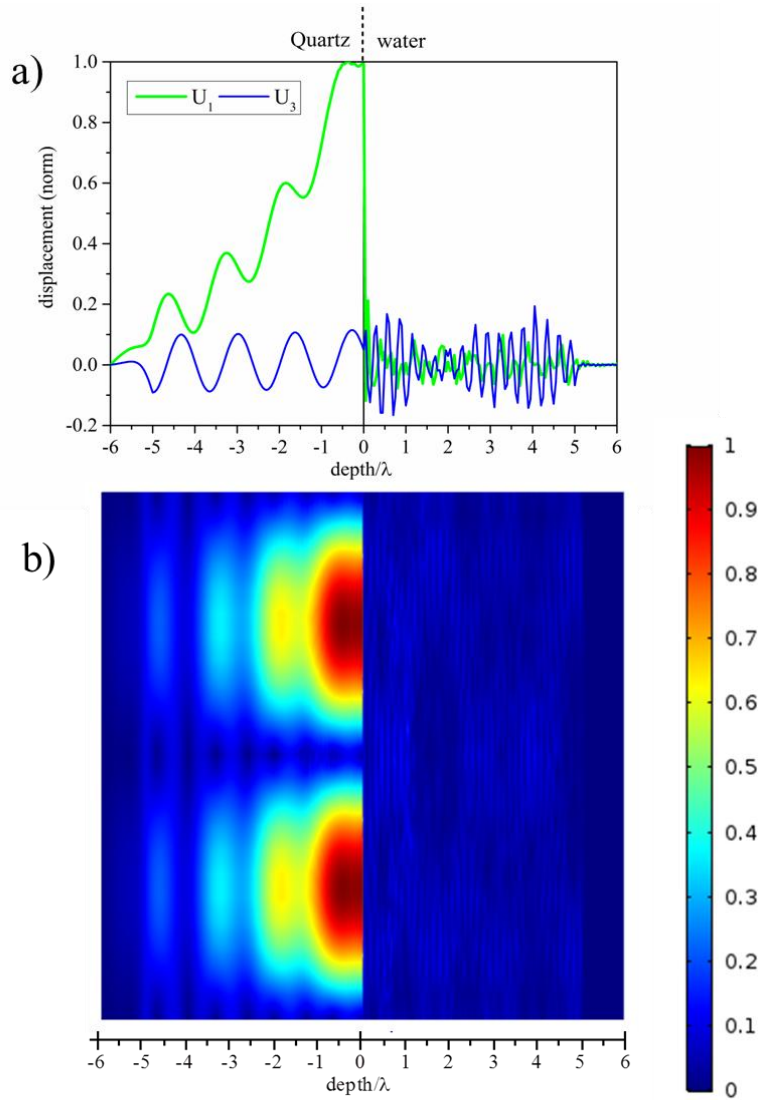

Figure 2.4: a) The HVPSAW displacement profile inside the ST-x quartz (from -120 to 0 μm abscissa values) and in water (from 0 to 120 μm abscissa value); b) the 2D representation of the total displacement confined inside the substrate.

In an attempt to design a sensor packaging able to protect the IDTs from the liquid environment and to confine the measurand in the device sensing area, the electroacoustic device can be integrated with different types of microchannels. Typically, the test cell that localises the liquid to the surface of the device consists in a pre-molded polydimethylsiloxane (PDMS) cell [31], or a SU8 cell [32] that is mechanically pressed against the surface of the sensor in order to avoid any contact between the IDTs and the liquid to be tested. The cell can be positioned in three different configurations: 1. in between the IDTs, as shown in figure 2.5a, to prevent the presence of liquid on the IDTs; 2. it can occupy the entire device surface if the IDTs are shielded by a proper layer, as shown in figure 2.5b, to avoid the conduction through the liquid between electrodes; 3. it can be positioned in between the IDTs and integrated with two air cavities that protect the IDTs: a vertical structure shaped as walls is positioned around the IDTs to separate them from the liquid filled container, as shown in figure 2.5c.

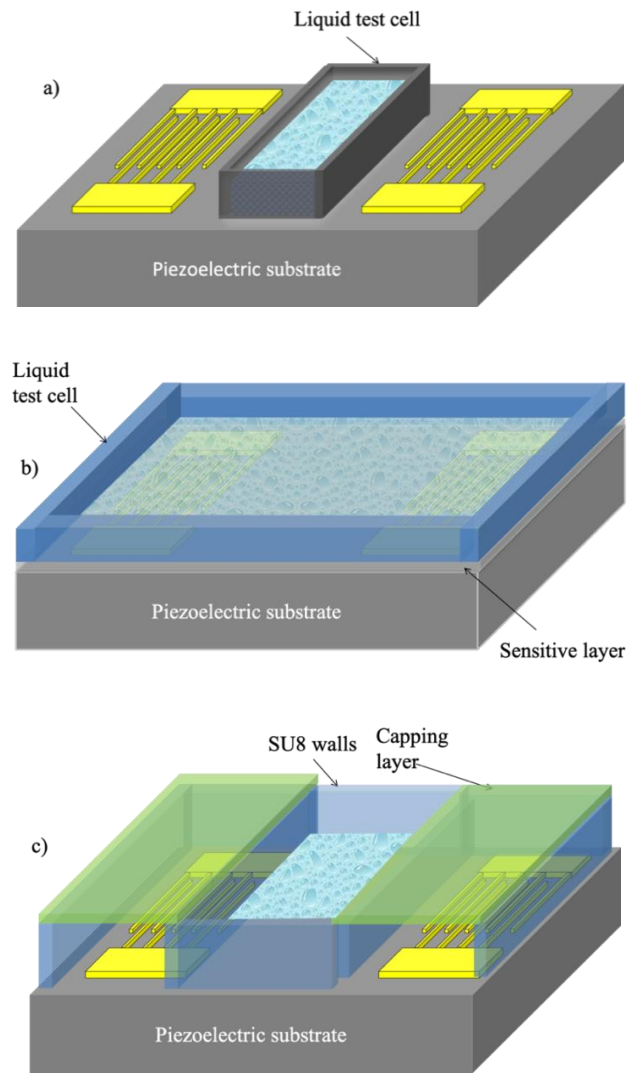

**Figure 2.5:** Schematic of the three sensor/liquid cell configurations: a) the liquid cell is located in between the IDTs; b) the liquid cell occupies the whole device surface; c) the IDTs are protected by a vertical structure from the liquid contact.

In the first and third configuration, the cell adds some disturbance effect on the propagation of the wave due to the mechanical stresses induced on the surface by the packaging, thus resulting in an additional improvement in damping losses. The second configuration ensures maximum sensitivity, as the physical size of the liquid cell includes the total wave propagation path, but requires a careful selection of the type and thickness of the *protective* surface layer in order to minimize potential perturbations to the device insertion loss, thermal stability, sensitivity, to cite just a few. The thin layer can even be a *sensitive* coating layer that exhibits high sensitivity toward a certain analyte but low sensitivity toward other species: thus the layer drives the sensor selectivity towards a specific target analyte. A liquid cell integrated with the sensor during the fabrication process is preferred as it reduces the devices complexity and enhance their performances.

References [33-35] provide a very useful list of substrates types and crystallographic orientations along which low-attenuated, strongly-coupled PSAW and HVPSAW travel (such as 64°YX

LiNbO$_3$, 36°YX LiTaO$_3$, quartz ST-X, LiNbO$_3$ with Euler angles (90° 90° 36°), (90° 90° 31°) LiTaO$_3$, and (0° 45° 90°) Li$_2$B$_4$O$_7$). For sake of completeness, the same references list other information that are fundamental for the design of the device parameters (operating frequency, IDTs centre-to-centre distance, number of IDTs finger pairs, IDTs aperture, etc.) such as the phase velocity, propagation loss, particle displacement components, electric-to-acoustic energy conversion efficiency, K$^2$, and power flow angle) of the SAW, PSAW and HVPSAW. Moreover, these references also describe the waves behaviour with increasing the thickness of a metallic layer covering the wave propagation path: depending on the layer thickness, the wave diffraction into the bulk can be prevented, as well as the transitions from HVPSAWs to higher order PSAWs modes, and from PSAWs to the SAWs can be observed. The piezoelectric substrates of the LGX-family group have been studied in references [36-38] for application to temperature stable, high coupling, low loss sensors for liquid environments. Several applications of PSAW sensors are presented in the available literature for the measurement of viscosity, electrical properties and mass loading of an adjacent liquid. In reference [39] a sensor for liquid viscosity and conductivity measurement is described that is based on a PSAW dual delay line on 41°-YX LiNbO$_3$ and covered by a SiO$_2$ protective layer. In reference [40] a PSAW sensor is implemented on 36°-YX LiTaO$_3$ substrate for methanol concentration measurement. In reference [41] the PSAW trapping efficiency of the free, metallized, and grating paths in YX-36° and YX-42° LiTaO$_3$ and YX-64° LiNbO$_3$ are compared at fundamental and harmonic frequencies. A high frequency (>500 MHz) PSAW sensor for liquid environment is designed that show high coupling, low loss, high operating frequency and high resistance to surface contamination.

In reference [42] the propagation of PSAW along bare 41° and 64°-Y LiNbO$_3$, and 36°-Y LiTaO$_3$ substrates is studied and compared with that along the same substrates covered by a thin sputtered glass films to increase the resistance to surface contamination. It was demonstrated experimentally and theoretically that the glass layer does not affect the wave loss but it lowers the temperature stability. Film thickness to wavelength ratio regions were found which represent a good trade-off between lowered temperature coefficient of frequency and a preserved high coupling factor. Sensors of liquid viscosity and mass loading have been demonstrated utilizing the present mode on 36°-YX LiTaO$_3$ [43].

One way to enhance the sensor sensitivity (i.e. the frequency change per unit incremental change of the measured quantity) is to raise the working frequency: this effect can be obtained by reducing the size of the IDTs metal strips or utilizing SAW devices based on high velocity acoustic wave modes. The HVPSAWs are attractive for this potential application as they travel at velocity close to the longitudinal bulk acoustic wave (BAW) velocity. In reference [44] the propagation of HVPSAWs in LiNbO$_3$ has been studied in the range of Euler angles (0°, 0° to 90°, 90°): the corresponding theoretical velocities are between 6700 m/s and 7400 m/s, about twice that of normal surface waves, but the K$^2$ varies between about 0.14% to 0.5%, much less than that of surface waves. In reference [45] some experiments show that the HVPSAW phase velocity in (0°, 124°, 50°) quartz reaches 6992 m/s and the propagation attenuation is as low as less than $1 \cdot 10^{-4}$ dB/$\lambda$, thus is suitable for liquid sensing applications. In reference [46] the propagation of HVPSAW along a semi-insulating Fe-doped GaN films grown on sapphire substrates is experimentally studied and the small propagation attenuation of the mode when travelling at a liquid/solid interface is demonstrated in glycerol solutions. In reference [47] the propagation loss due to bulk wave radiation of a HVPSAW is reduced by loading the 36°YX -LiNbO$_3$ substrate with a dielectric amorphous AlN thin film with a higher velocity than the substrate. The amorphous AlN layer plays the double role to protect the IDTs patterned onto the substrate surface and to enhance the device performances. Table 2.1 list some examples of PSAW-based sensors and just two examples related to the HVPSAW: it is evident that, despite the long-time-recognized suitability of the HVPSAWs for liquid sensing applications [48] there is a lack of experimental validations.

Table 2.1. Some practical examples of SAW sensors for application to liquid environment test.

| Wave type | substrate | layer | application | reference |
|---|---|---|---|---|
| PSAW | 41°-YX LiNbO$_3$ | SiO$_2$ | liquid viscosity and conductivity | [39] |
| PSAW | 36°-YX LiTaO$_3$ | -- | methanol concentration | [40] |
| PSAW | 36°-YX LiTaO$_3$ | -- | liquid viscosity and mass loading | [43] |
| HVPSAW | Fe-doped GaN/sapphire | -- | Viscosity-density product | [46] |
| PSAW | 36°YX-LiTaO3 | poly(methyl methacrylate) (PMMA) or cyanoethylcellulose (CEC) | rabbit anti-goat IgG | [49] |
| PSAW | 64° YX-LiNbO$_3$ | 1,10-phenanthroline | Heavy metal compounds: PbNO3 and CdNO3 | [50] |
| PSAW | 36° YX-LiTaO3 | -- | tiny particles mixed with distilled water | [51] |
| PSAW | 36°YX-LiTaO3 | -- | detection and discrimination of various detergents | [52] |

| PSAW | 36°YX-LiTaO3 | parylene thin film | Urea biosensing | [53] |
| HVPSAW | Quartz (0° 124° 50°) | -- | Water loading | [48] |

### 3. Love wave sensors

Love waves are a type of surface acoustic waves characterized by a shear horizontal particle displacement component $U_2$ dominant over the vertical and longitudinal ones ($U_2 >> U_3, U_1$). The propagation of the Love waves is excited and revealed by means of a couple of IDTs, as for the SAW-based devices. Due to the in-plane polarization, the Love waves, as well as the PSAW and HVPSAWs, are suitable to travel at a surface contacting a liquid environment. In the most general sense, Love waves propagate along the surface of a piezoelectric half-space covered by a thin layer: the substrate is responsible for the excitation of a surface skimming *bulk* wave (SSBW) that propagates below the substrate surface; the thin over-layer traps the acoustic energy and slows down the wave propagation velocity, thus reducing the loss from radiation into the bulk. As a result, the SSBW is converted into a shear *surface* wave, the Love wave. Figure 3.1 shows the field profile of the SSBW and of the Love wave travelling at the surface of a *bare* ST quartz substrate, and of the Love wave tavelling at the same substrate covered by a thin $SiO_2$ trapping layer. As it can be seen, the strain associated to the SSBW penetrates deep within the bulk of the bare quartz substrate; the strain associated to the Love wave remains close to the surface of the $SiO_2$/quartz substrate.

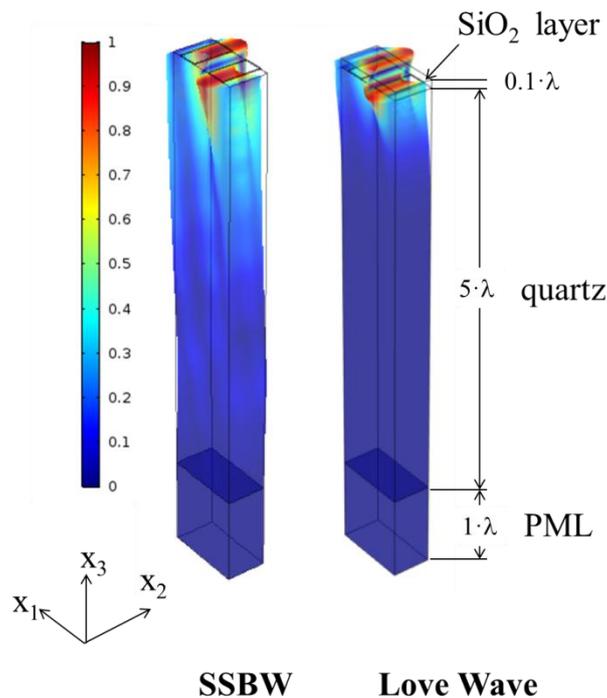

**SSBW     Love Wave**

Figure 3.1: The field profile of the SSBW and Love wave travelling along the *bare* and the SiO₂ film covered ST-90°x quartz substrate in air; x₁ is the wave propagation direction and x₃ is the normal to the substrate surface.

The number of Love modes (LMs) that can propagate in the layer/substrate medium depends on the layer thickness, but he essential condition for the propagation of the Love waves is that shear bulk wave velocity (SHBAW) of the halfspace, $v_{sub}^{SHBAW}$ , is larger than the shear bulk wave velocity of the layer, $v_{layer}^{SHBAW}$ , as the velocity of the Love wave lies in between the $v_{sub}^{SHBAW}$ and the $v_{layer}^{SHBAW}$ [54]. Higher order LMs develop at their respective cut-off frequencies, which are related to the thickness of the layer: they are dispersive as their velocity depends on the layer thickness, other than on the substrate and the layer's material properties. As the LMs acoustic energy is mostly concentrated inside the guiding layer, the LM-based devices show good performances in terms of sensitivity to any disturbance loading the surface of the guiding layer. A comprehensive review of the Love wave sensors can be found in Reference [55]. As an example, Figure 3.2 shows the phase velocity dispersion curves of the first five modes (LM1, LM2, LM3, LM4 and LM5) travelling along the ST 90°-x quartz half-space covered by a SiO₂ film. The picture also shows the $v_{sub}^{SHBAW}$ and the $v_{layer}^{SHBAW}$ . The velocity values were numerically calculated by using the McGill software [56].

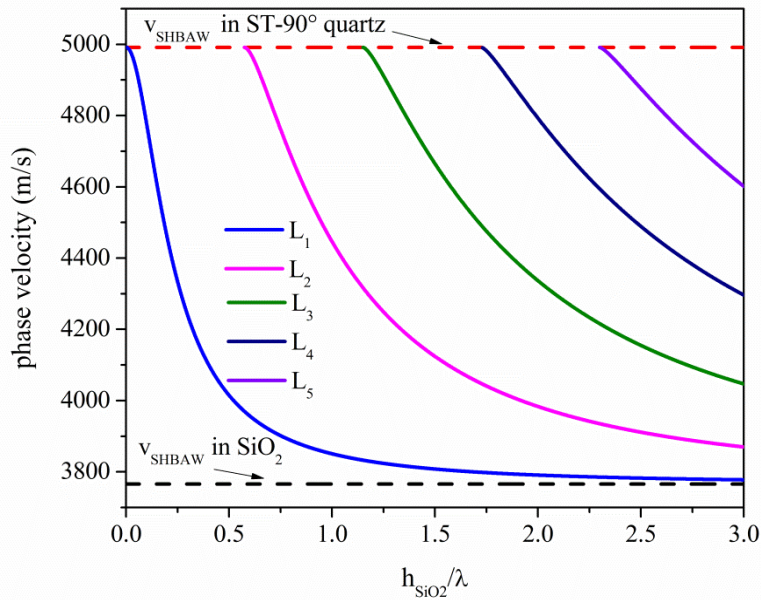

Figure 3.2: The phase velocity vs the layer normalized thickness of the first four LMs travelling in ST 90°-x quartz/SiO₂ substrate.

When the guiding layer is very thin, the velocity of the LM1 tends to the half-space SHBAW velocity; with increasing the layer thickness, the velocity of both the fundamental and higher order modes asymptotically reaches the layer SHBAW velocity. Love waves vanish if the frequency is lower than the cut-off frequency.

Figure 3.3a shows the delay line on top of the quartz half-space with the two IDTs located at a distance equal to $3 \cdot \lambda$ ($\lambda = 20$ μm); the depth of the substrate is $10 \cdot \lambda$. The figure represents the total displacement 1ns after the electric signal is applied at the transmitting IDT. Figure 3.3b shows the time evolution of the SSBW and LM1 total displacement propagating at the surface of the ST 90°-x quartz half-space, bare and covered by a SiO$_2$ trapping layer, 2 μm thick. The total displacement of the SSBW and LM1 was calculated by 3D COMSOL simulation: the time domain analysis was carried for 20 ns and the total displacement of the propagating medium was recorded at an interval of 1 ns. The transmitting bidirectional IDT launches two waves in opposite directions, as indicated by the arrows in figure 3.3b; the signal applied at the transmitting IDT was a 10 V peak-to-peak sinusoidal signal at 231 and 217 MHz, for SSBW and LM1. The plot clearly shows that the acoustical displacement propagates into the depth of the substrate for the SSBW while it is more confined to the surface for the LM1.

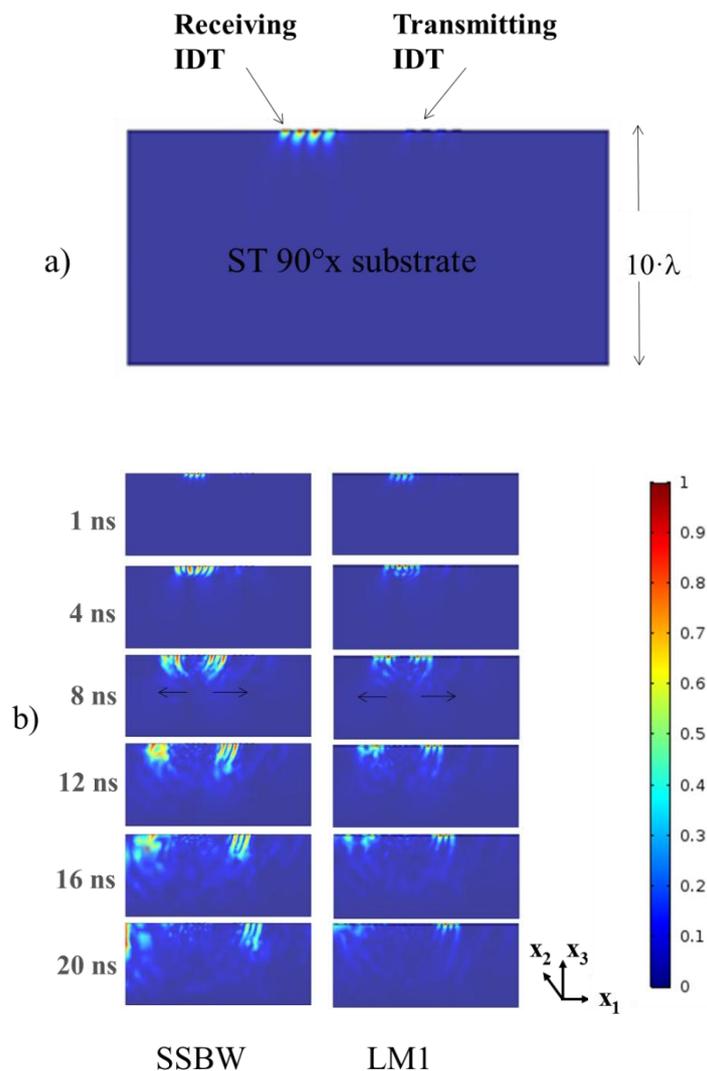

Figure 3.3. a) The transmitting and receiving IDTs positioned onto the quartz substrate; b) The time evolution (1 to 20 ns from the wave excitation) of the total displacement of SSBW and LM1; the two waves propagate along the ST 90°-x quartz half-space, bare and covered by a SiO$_2$ guiding layer, 2 μm thick, being λ = 20 μm.

The time and frequency domain response analysis of a delay line based on the SSBW and LM1 propagation was performed by 3D COMSOL simulation assuming a pair of Al IDTs, 0.1 μm thick, positioned onto the quartz surface (see figure 3.4). The distance between the transmitting and receiving IDT was set equal to 3·λ (active gap region) and the fingers overlap W =1·λ; the substrate propagation loss was not accounted in the calculation. The IDT number of finger pairs N was assumed to be equal to 2 for both the SSBW and LM1 devices. The electrical voltage at the receiver electrode was recorded for 30 ns in the time domain analysis, as shown in figure 3.4a where the ratio $V_{out}/V_{in}$ of the voltage at the receiver and transmitter IDT in time domain is shown. The insertion loss of the delay line is calculated by applying a unit impulse at the input IDT: the Fourier transformation of the device impulse response allowed the calculation of the scattering parameter of the SSBW and LM1 delay line, $S_{21} = 20 \cdot \log(\text{Re}[V_{out}/V_{in}])$, shown in figure 3.4 b. For more information on the scattering parameters, please refer to the textbook [21]. $V_{in}$ and $V_{out}$ are the voltages at the alternate fingers of the transmitting and receiving IDTs respectively, while the remaining IDTs fingers are grounded.

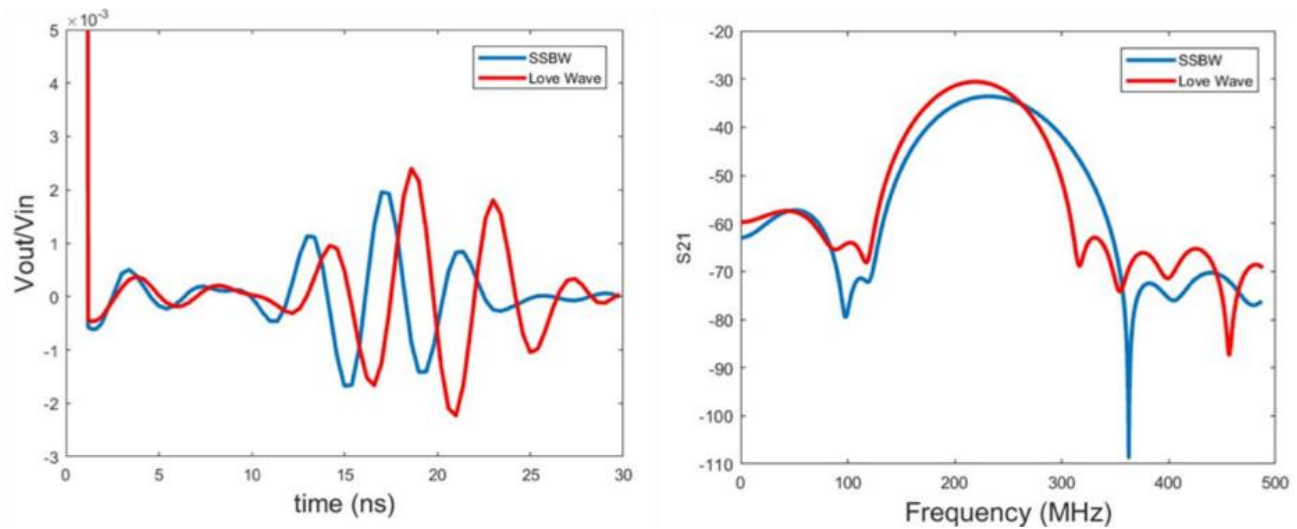

**Figure 3.4** a) Calculated ratio $V_{out}/V_{in}$ of the voltage at the receiver and transmitter IDT in time domain; b) calculated scattering parameter $S_{12}$ vs frequency, for the SSBW- and LM1-based delay line onto the ST-90°x quartz substrate, bare and covered by a SiO$_2$ guiding layer, 2 μm thick.

The voltage $V_{out}$ at the output IDT starts to rise in about 11 and 12.5 ns which corresponds to a SSBW and LM1 group velocities of 3636 and 3250 m/s, respectively.

The presence of the guiding layer is fundamental to trap the Love wave energy, and only a well-defined thickness guarantees enhanced device performances (such as minimum delay line insertion loss and maximum gravimetric sensitivity). Figure 3.5 shows the $K^2$ dispersion curve (the black curve) and the derivative of the group and phase velocity respect to the normalized thickness of the SiO$_2$ layer, $\dfrac{\partial v_{gr}}{\partial\left(h_{SiO2}/\lambda\right)}$ and $\dfrac{\partial v_{ph}}{\partial\left(h_{SiO2}/\lambda\right)}$, vs $h_{SiO2}/\lambda$ curves for the LM1 travelling along the ST 90°-x/SiO$_2$ substrate. The $h_{SiO2}/\lambda = 0.074$ corresponds to the maximum $K^2 = 0.227\%$; $h_{SiO2}/\lambda = 0.115$ and $0.08$ correspond to the maximum mass sensitivity of the phase and group velocity of the LM1, as the derivative of the $v_{gr}$ and $v_{ph}$ is proportional to the gravimetric sensitivity $S_{grav} = \left(\Delta v/v_0\right)/(\rho \cdot h_{SiO2})$, where $\rho$ and $h_{SiO2}$ are the layer mass density and thickness, $\Delta v = v - v_0$, $v_0$ and $v$ the wave velocity along the bare and covered half-space [57]. The $K^2$ values calculated at the abscissa values corresponding to the $v_{gr}$ and $v_{ph}$ maximum sensitivity are quite similar (0.230% and 0.204%), while the expected $v_{gr}$ sensitivity is predicted to be about twice that of the $v_{ph}$.

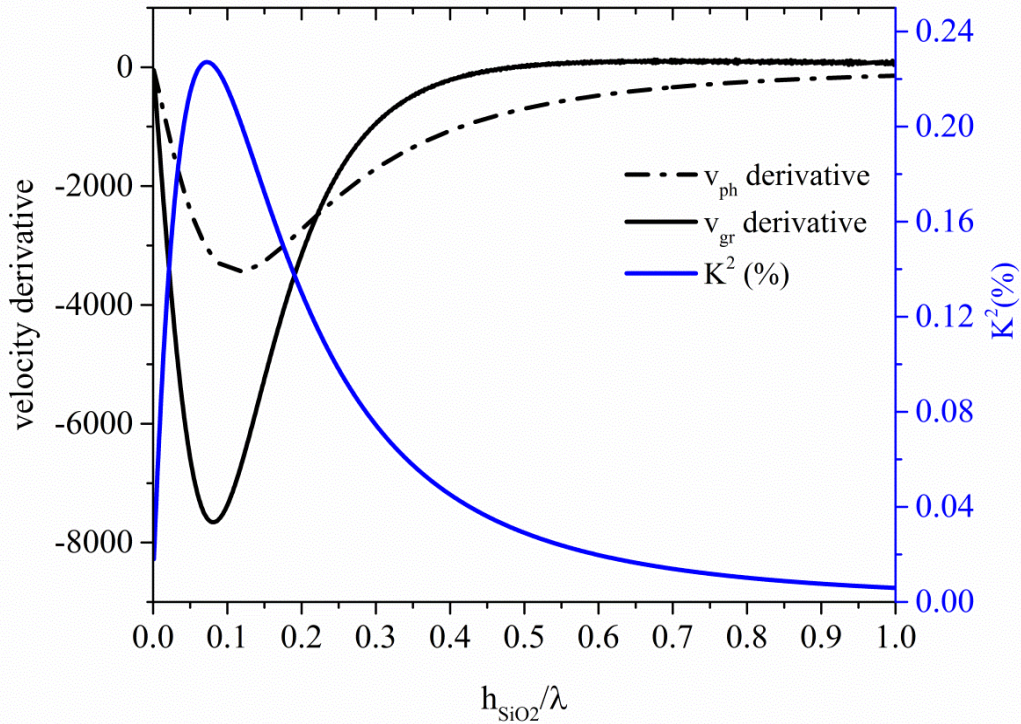

Figure 3.5: The $K^2$ and the derivative of the phase and group velocity vs the normalized SiO$_2$ layer thickness, $h_{SiO2}/\lambda$, of the fundamental LM1 travelling along the ST 90°-x/SiO$_2$ substrate.

Figure 3.6 shows the derivative of the $v_{ph}$ and $v_{gr}$ of the first five LMs in ST 90°-x quartz/SiO$_2$ with respect to the layer normalized thickness. As it can be seen, the magnitude of the gravimetric sensitivity $S_{grav}$ increases with increasing the layer thickness and reaches a peak after which, with increasing the guiding layer thickness, it decreases. The peak of the $v_{ph}$ sensitivity decreases with increasing the mode order and the highest value corresponds to the LM1 mode; the $v_{gr}$ mass sensitivity increases rapidly with the layer thickness, and it can be larger than the former as its peak can be sharper.

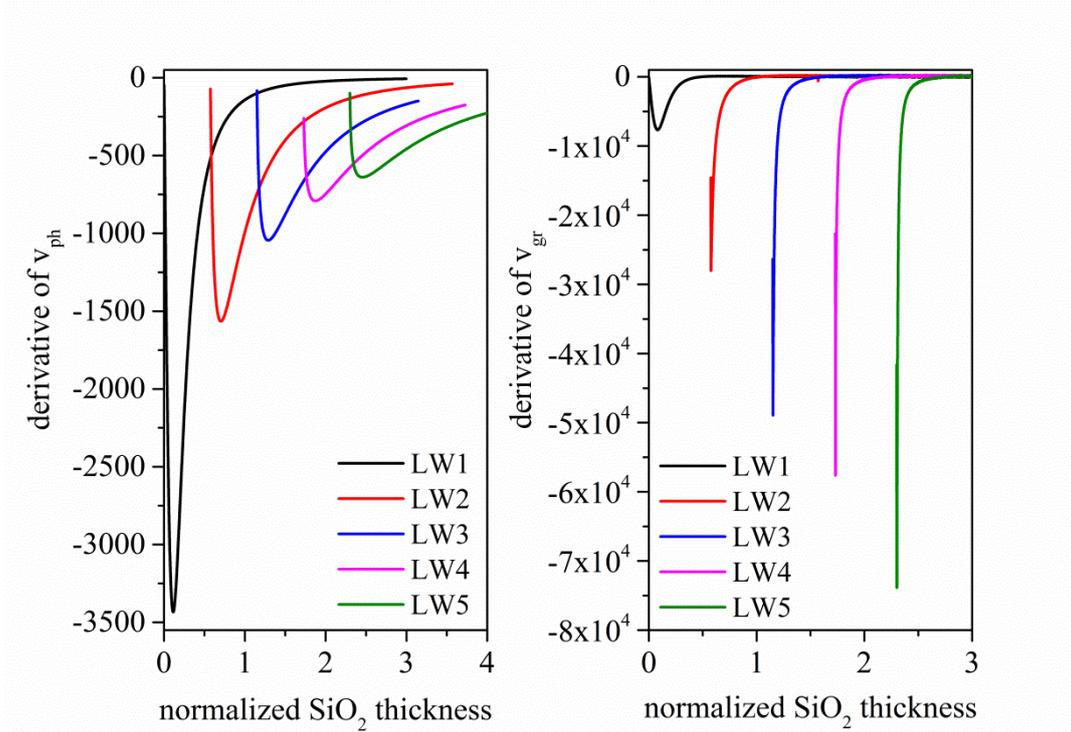

Figure 3.6: The derivative of the phase and group velocity vs the normalized layer thickness for the first five LMs travelling along the ST 90°-x quartz/SiO₂ substrate.

Both the LMs group and phase velocity can represent a sensor response [58]: the phase velocity can be experimentally estimated by measuring the operating frequency $f = v_{ph}/\lambda$ of the sensing device at the minimum insertion loss of the scattering parameter S₁₂. The group velocity can be estimated by measuring the group time delay $\tau = L/v_{gr}$ of the sensing device at the minimum insertion loss of the scattering parameter S₁₂ in the time domain, being L the acoustic wave delay path (the IDTs centre-to-centre distance)

In the most general cases the LMs devices consist of a semi-infinite piezoelectric substrate (for example 41°YX LiNbO₃, 36°YX LiTaO₃ and ST-90°X quartz) [59] covered by a thin slowing layer (for example ZnO, Au, PMMA or SiO₂) which traps the propagating wave to the surface of the substrate. The IDTs can be located only onto the piezoelectric substrate surface, under the overlayer, and thus they are isolate from the liquid environment. Table 3.1 lists some practical examples of LM sensors for application to liquid environment.

Table 3.1. Some practical examples of LM sensors for application to liquid environment.

| substrate | layer | Application | reference |
|-----------|-------|-------------|-----------|
| ST-90° | SiO₂ | mass sensing in liquids | [60] |
| quartz | | | |
| ST-90° quartz | PMMA | mass sensing in liquids | [61] |
| LiTaO$_3$ | SiO$_2$, ZnO, gold, SU-8, and parylene-C | Comparison of electromechanical coupling coefficient, displacement profile and mass sensitivity | [62] |
| ST-90° quartz | ZnO | liquid viscosity and conductivity | [63] |
| ST-90° quartz | PMMA | detection of high molecular weight targets in liquid samples | [64] |
| 36°-YX LiTaO$_3$ | ZnO | methanol in water | [65] |
| 36°-YX LiTaO$_3$ | ZnO | antibody–antigen immunoreactions in aqueous solutions | [66] |

The LMs also propagate along a non-piezoelectric halfspace (such as Si, glass, BN, a-SiC,...) covered by a piezoelectric layer (such as c-axis tilted ZnO or AlN) [58, 67-69].

For example, when the hexagonal ZnO film has its c-axis parallel to the substrate free surface, it is effective in the electroacoustic transduction of LMs in glass/ZnO substrate; when the c-axis is tilted at an angle μ with respect to the normal to the substrate surface, for wave propagation along the <100> direction, two types of surface modes propagate, the LM with predominant shear horizontal polarization, and the Rayleigh-like, with a prevailing sagittal polarization. Both modes are coupled to the electric field via the effective piezoelectric constants of the ZnO film. The LM and the SAW play different roles onto the same sensing platform: the former is suitable for liquid environment characterization, while the latter is suitable for mixing and pumping small liquid volumes. LM sensors implemented on silicon or glass substrate materials offer the great advantage of the sensor's integration with the surrounding electronic circuits.

For biosensing applications, a sensitive layer can be positioned along the acoustic wave propagation path: in this case the thickness of the sensing membrane must be properly designed in order not to perturb the wave propagation characteristics. Figure 3.7 shows the schematic representation of Love wave sensor. The membrane can cover only the path in between the IDTs or the entire wave path if the IDTs are buried under the guiding layer. In the latter case, the trapping layer must satisfy also the requirement of good chemical and mechanical resistance [70] as it has the additional role of shielding the IDTs.

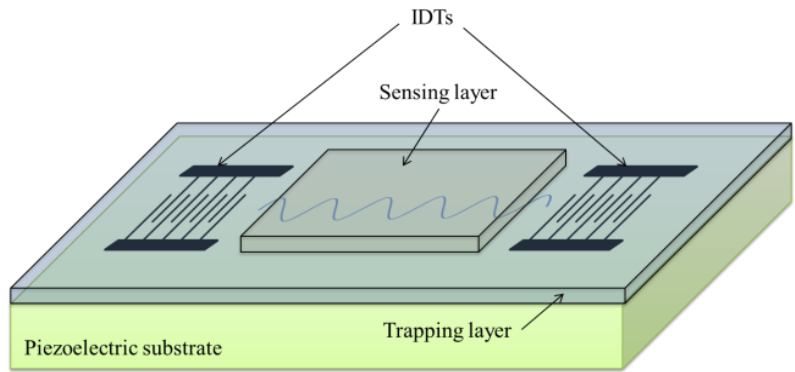

Figure 3.7: The schematic of the Love wave device.

The Love wave trapping layer can be by a thin polymeric film such as polymethylmethacrylate (PMMA) [71], polyimide, SU-8 or parylene C. In reference [61] the adsorption of h-IgG on the polymer surface was investigated by using a 1.23 µm thick PMMA guiding layer onto quartz Love wave device, and finally the potential of the device as a biosensor was investigated by detecting the binding of anti-IgG.

The LM sensor based on a piezoelectric layer/non piezoelectric substrate has a remarkable advantage over their counterpart based on piezoelectric half-space/non-piezoelectric layer, as well as over PSAW and HVPSAW-based sensors: four coupling configurations can be investigated to enhance the $K^2$ and to take advantage of the protecting role of the guiding layer if the IDTs are buried under it. The IDTs can be positioned at the layer/substrate interface (substrate/transducer/film, STF) or at the layer free surface (substrate/film/transducer, SFT), with or without a floating metal layer onto the opposite surface of the layer (substrate/transducer/film/metal, STFM, or substrate/metal/film/transducer, SMFT), as shown in figure 3.8.

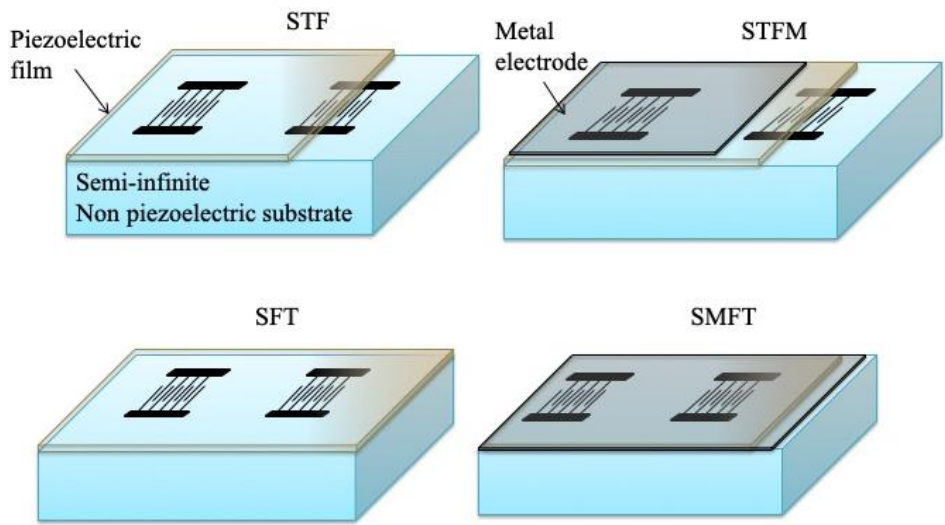

Figure 3.8. The four coupling configurations for the non-piezoelectric half-space/piezoelectric layer structure [58].

The $K^2$ of the LM device is affected by the mode order, the crystallographic orientation of both the halfspace and layer, the layer thickness, and also the coupling configuration (the electrical boundary conditions). As an example, figure 3.9 shows the $K^2$ dispersion curves for the first Love mode (LM1) in ZnO/glass for the four coupling configurations and different ZnO c-axis tilt angle (from 10° to 90°) [58].

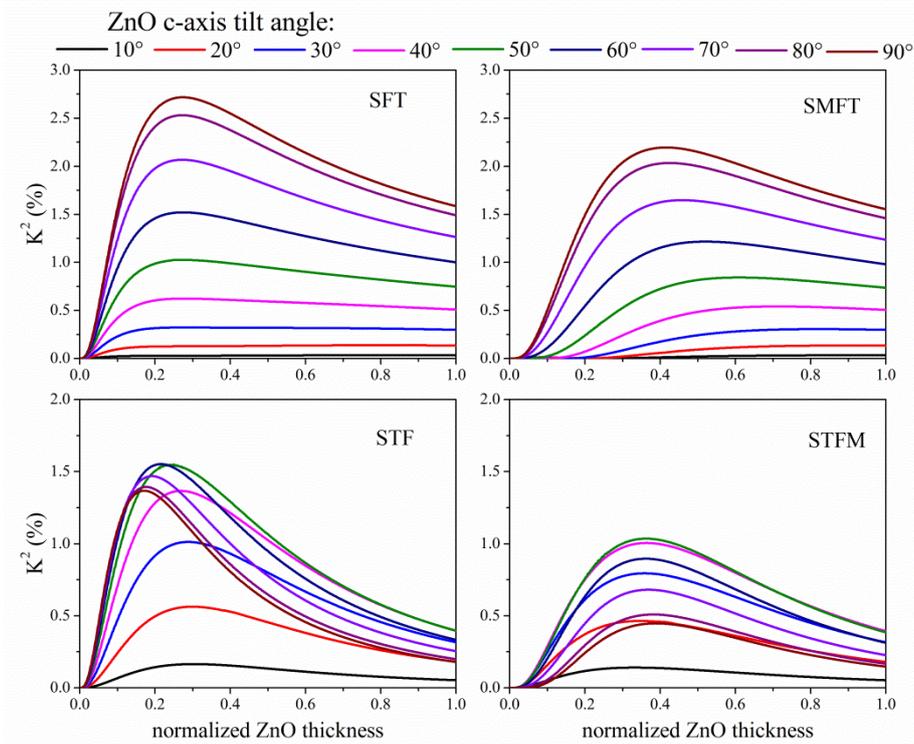

Figure 3.9: The $K^2$ dispersion curves for the first Love mode in ZnO/glass for the four coupling configurations and different ZnO c-axis tilt angle (from 10° to 90°). The colour of each curve represents a different c-axis tilt angle.

The SFT (SMFT) configuration reaches the highest $K^2$ values for h/λ ~ 0.3 (0.4) for large tilt angles; the STF and STFM configurations reaches their maximum $K^2$ value (~1.6 and 1.1 %) for 50° tilt angle.

As an example, figure 3.10 shows the $K^2$ dispersion curves for the first four Love modes (LM1, LM2, LM3 and LM4) in 90°-tilted c-axis ZnO/wBN, for the four coupling configurations.

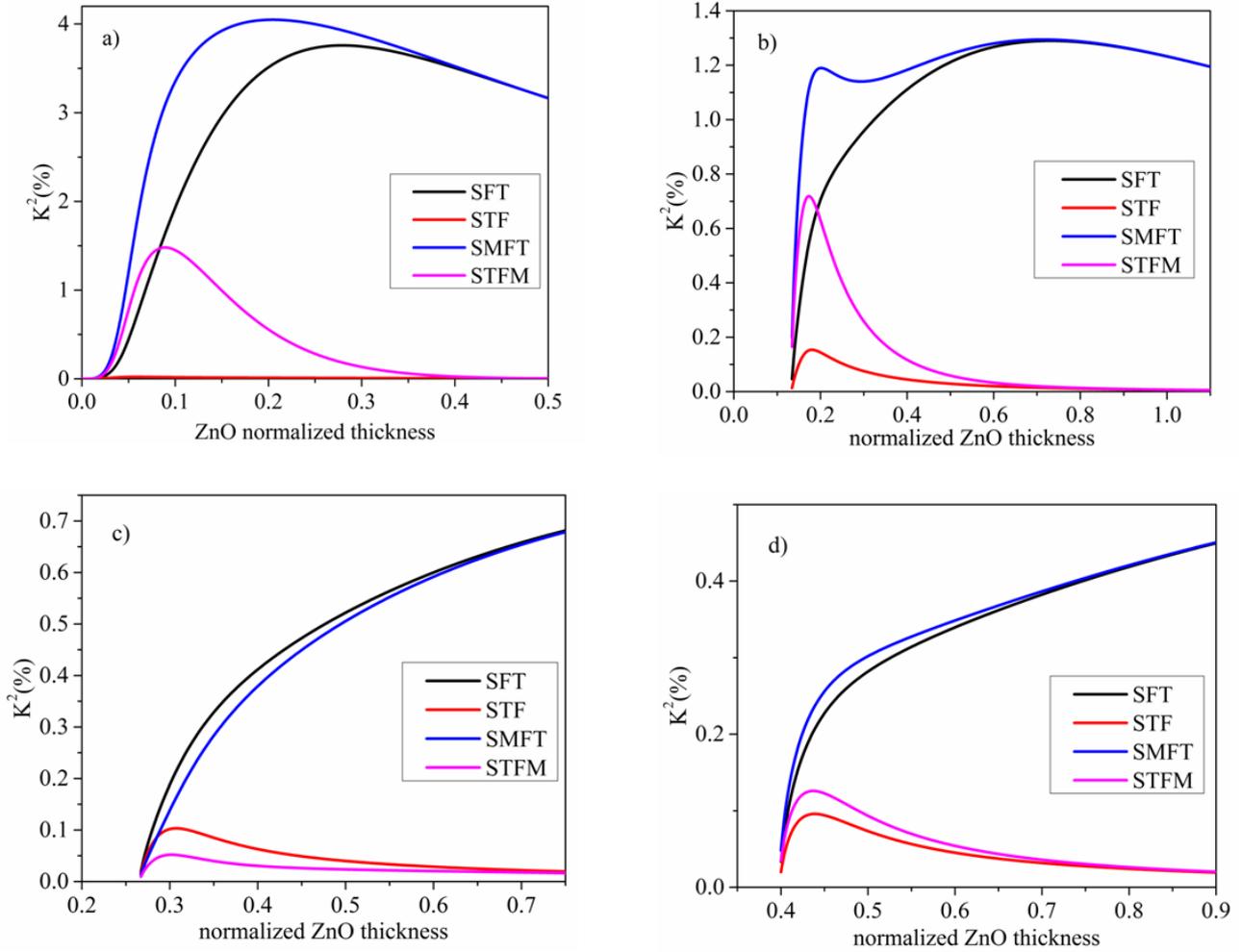

Figure 3.10: The K$^2$ dispersion curves for the a) LM1, b) LM2, c) LM3 and d) LM4 modes in 90°-tilted c-axis ZnO/wBN, for the four coupling configurations [68].

With increasing the Love mode order, ever decreasing K$^2$ values can be reached by the 4 coupling configurations. The remarkable advantage of the LM-based sensors fabricated onto silicon is the possibility to integrate the sensor with other devices.

## 4. Shear Horizontal Acoustic Plate Mode sensors

Shear Horizontal Acoustic Plate Modes (SHAPMs) are waveguide modes that propagate in finite thickness plates with energy distributed throughout the bulk of the waveguide. The SHAPMs are shear horizontally polarized (U$_1$, U$_3$ = 0), hence the absence of the out-of-plane displacement component allows each mode to propagate in contact with a liquid without coupling excessive amounts of acoustic energy into the liquid. 3D eigenfrequency FEM analysis was performed using COMSOL Multiphysics® Version 5.2 to explore the field shape of the SH0, SH1, SH2 and SH3 travelling along the along a GaPO$_4$ piezoelectric plate, 150 µm thick, with Euler angles (0° 5° 90°) and thickness to wavelength ration h/λ = 0.6, as shown in figure 4.1. The GaPO$_4$ materials constants are those provided by Piezocryst Advanced Sensorics GmbH, which is an European GaPO$_4$ wafers supplier [72, 73].

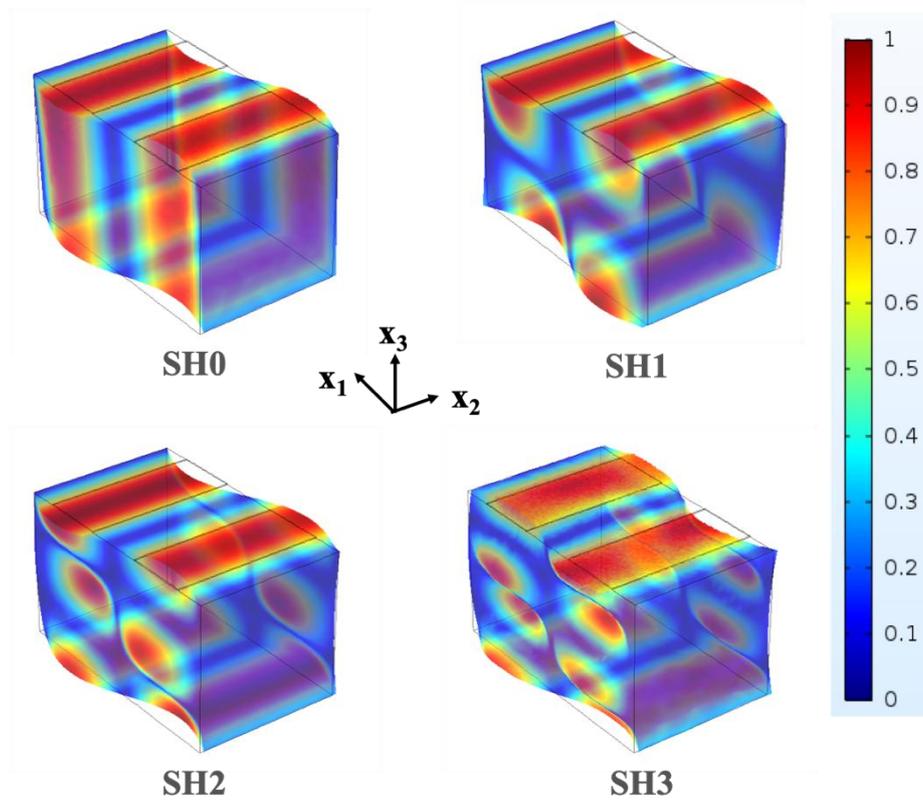

Figure 4.1: The field profile of the first four SHAPMs in GaPO$_4$ piezoelectric plate in contact with air.

The SHAPMs-based device employs input and output IDTs to launch and receive the acoustic wave, like for the SAWs–based devices. These modes energy is distributed between the two plate surfaces as for a standing wave in a BAW sensor but the SHAPMs travel along the plate like a SAW. The continuous exchange of energy between the two plate surfaces allows the signal between the two IDTs to be affected by any changes of the surrounding environment the opposite plate sides undergoes. For liquid sensing applications the plate itself can be employed as a physical barrier between the electronics and the liquid environment to be sensed. The IDTs may be placed onto the surface opposite to the one in contact with the liquid solution, as shown in figure 4.2: the IDTs are naturally isolated from the (potentially corrosive) liquid environment without adding any protective layer to the device surface, as for the SAW-based sensors, thus taking advantage of the entire sensor surface to maximize the interaction of the wave with the analyte. A metal film can be placed between the input and output IDTs to cancel any direct electromagnetic feedthrough. A sensing membrane may be attached to the upper side of the plate that is selectively sensitive to a specific measurand contained into the test liquid solution contacting the sensor. Any interaction (mechanical and/or electrical) between the measurand and the sensing membrane will cause a shift in the attenuation and/or velocity of the wave, which represents to the sensor response.

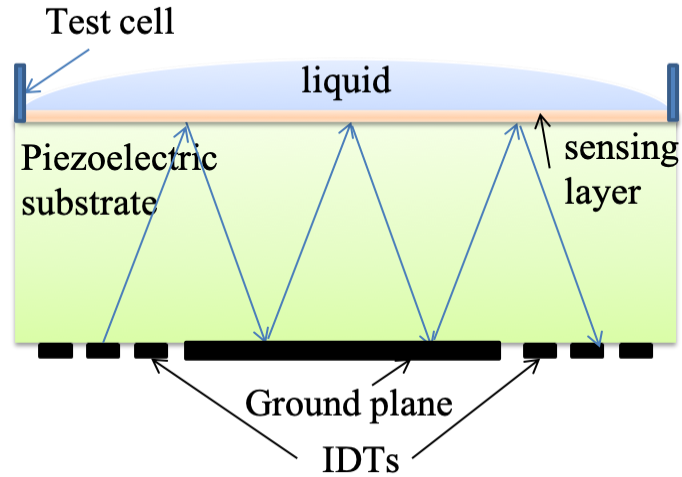

Figure 4.2: The schematic of the SHAPMs sensor including a sensitive membrane [74].

The number of modes that propagate along the plate is dependent on the normalized thickness h/$\lambda$ of the plate; the modes are excited at frequency $f_n=v_n/\lambda$ where $v_n$ is the velocity of the n-th mode corresponding to the selected h value. As an example, Figure 4.3 shows the phase velocity dispersion curves of the first six SHAPMs travelling along an y-rotated GaPO$_4$ plate with Euler angles (0° 1° 90°); the data were obtained by using the McGill software [56]. As it can be seen, the fundamental mode, SH0, is a low-dispersive symmetric mode that travel at velocity equal to the transverse BAW velocity. The higher order modes can be symmetric and anti-symmetric: they are highly dispersive and their velocity asymptotically reaches the shear BAW velocity with increasing the plate thickness; they have a cut off thickness: below the cut-off frequency, the mode becomes evanescent, i.e., the wavenumber is imaginary. Higher order modes can reach very high velocity as near the cut off the slope of the dispersion curves is near to be infinite.

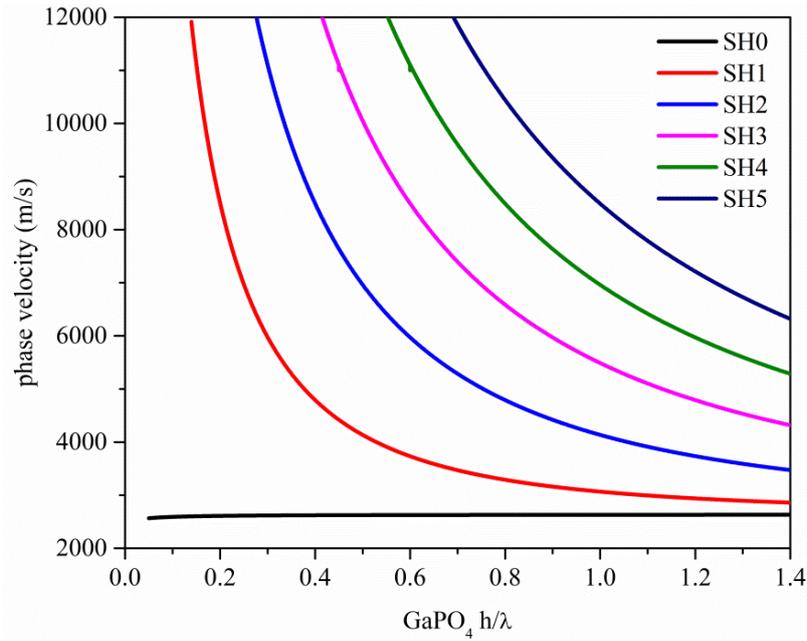

Figure 4.3: The phase velocity dispersion curves of the first six SHAPMs in $GaPO_4$ plate ($0°$ $1°$ $90°$) [74].

SHAPMs have maximum displacements that occur on the top and bottom surfaces of the plate, with sinusoidal variation between the two plate sides. The field profile of the first four SHAPMs (SH0, SH1, SH2, and SH3) in ZnO ($0°$ $90°$ $0°$) with $h/\lambda$=0.5 are shown in figure 4.4. SHAPMs are divided into symmetric and anti-symmetric modes: for each mode the maximun displacement occurs on the top and bottom surfaces of the plate, allowing the use of either side of the plate for liquid sensing applications; the number of zeros is equal to the order of the mode. The fundamental symmetric mode (SH0 in figure 4.4a) differs from the others in that the acoustic field is uniformly distributed along the plate depth.

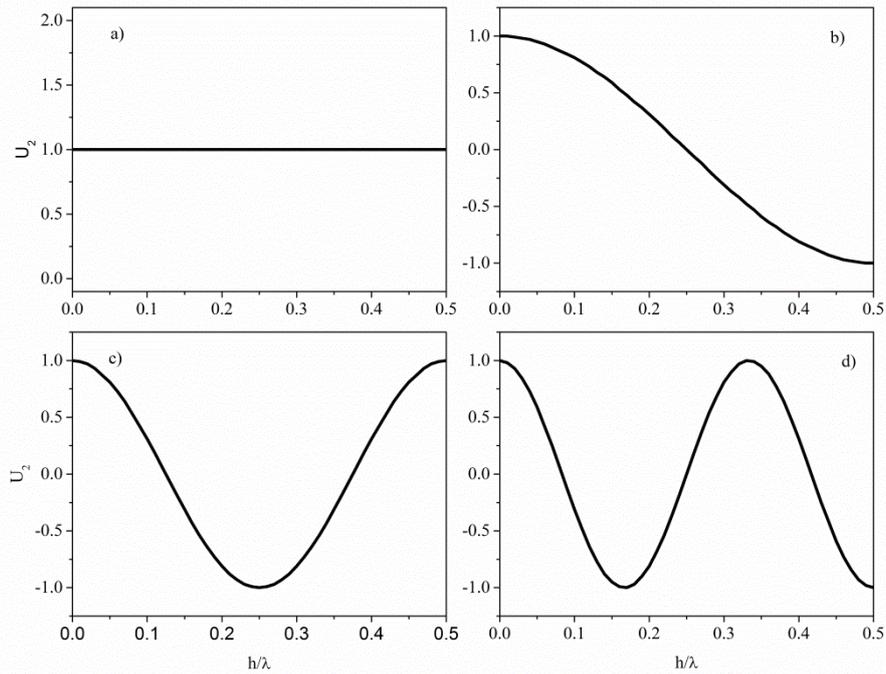

Figure 4.4: Cross-sectional displacement profiles for the four lowest-order SHAPMs in ZnO (0° 90° 0°) with normalized thickness h/λ=0.5 [74].

Excitation of shear plate modes, showing dominant shear horizontal polarization, can be accomplished by tilting the c-axis away from the vertical by an angle α. Pure SHAPMs exist on 90°-x propagating rotated y-cut of trigonal class 32 group crystals, which include the GaPO$_4$ and the quartz crystals [75, 76], and in x-propagating rotated y-cut hexagonal plates, such as AlN or ZnO. Figure 4.5 shows the rotated crystallographic system of the piezoelectric plate.

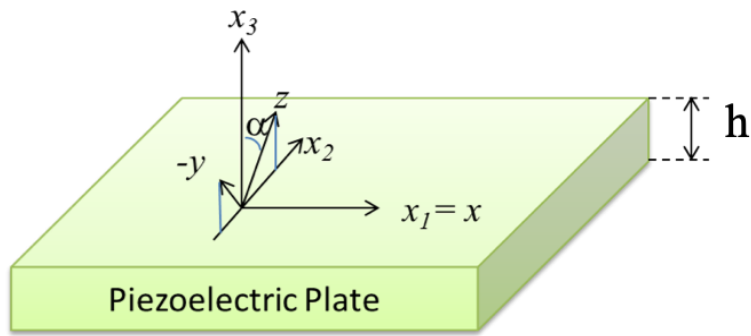

Figure 4.5: The rotated crystallographic system of the piezoelectric plate: x, y, and z represent the crystallographic axis system; $x_3$ is the plate normal; the propagation direction of the pure SHAPM is equal to $x_2$ or $x_1$ for trigonal or hexagonal crystals [74].

The tilt angle α, as well the plate thickness, affects the phase velocity and hence the $K^2$. As an example, figures 4.6a-b show the $K^2$ of the two coupling configurations, ST and MST, on ZnO vs the normalized plate thickness, being the tilt angle α the running parameter: the data were calculated with McGill software [56]. ST stands for substrate/transducer and is referred to the normal case where the IDTs are placed on one plate side, while MST stands for metal/substrate/transducer and is referred to the previous configuration with the opposite surface covered by a floating mass-less, infinitesimally thin metal layer.

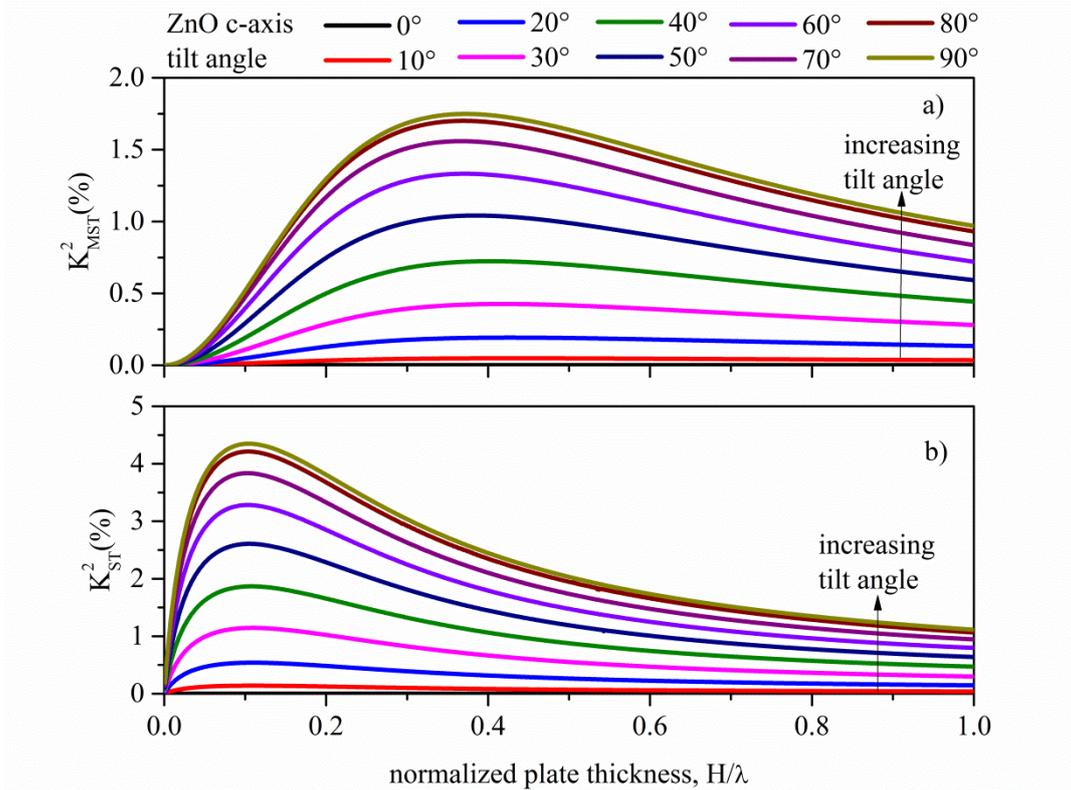

Figure 4.6: The $K^2$ of the a) MST and b) ST coupling configuration on c-axis tilted ZnO plate vs the normalized plate thickness, being the ZnO c-axis tilted angle α the running parameter.

The viscosity sensitivity of the SHAPM sensors as well as the relative surface displacement (and particle velocity) increase with increasing mode order [1, 77, 78]. Figure 4.7a, b and c show the attenuation of the SH1, SH2 and SH3 modes vs the frequency thickness product for a glass plate immersed in ethylic alcohol ($\rho$ = 790 kg/m$^3$, $v_l$ = 1238 m/s, dynamic viscosity = 1.2·10$^{-3}$ Ns/m$^2$, kinematic viscosity = 1.52·10$^{-6}$ m$^2$/s), benzene ($\rho$ = 881 kg/m$^3$, $v_l$ = 1117 m/s, dynamic viscosity = 0.65·10$^{-3}$ Ns/m$^2$, kinematic viscosity =7.38·10$^{-7}$ m$^2$/s) and kerosene ($\rho$ = 822 kg/m$^3$, $v_l$ = 1319 m/s, dynamic viscosity = 1.5·10$^{-3}$ Ns/m$^2$, kinematic viscosity =1.82·10$^{-6}$ m$^2$/s). The modes are sensitive to the viscosity of the liquids even when the mass density $\rho$ and/or the velocity $v_l$ are quite similar.

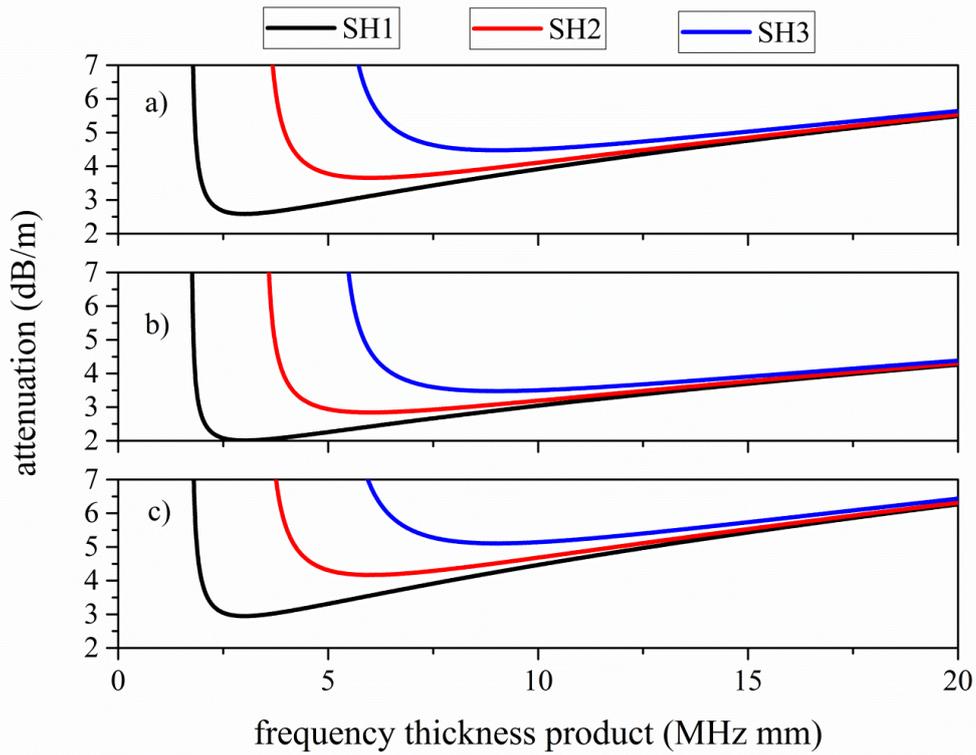

Figure 4.7: The attenuation of the SH1, SH2 and SH3 modes vs the frequency thickness product for a glass plate immersed in a) ethylic alcohol, b) benzene and c) kerosene.

The sensitivity also increases as the device is thinned: the lower limit of the plate thickness is limited by production processes and plate fragility. Figure 4.8a and b show the attenuation and the phase velocity of the SH1 mode vs frequency for three different thicknesses (1, 1.2 and 1.4 mm) for a glass plate immersed in kerosene: the curves move toward higher attenuation and velocity values with decreasing plate thickness. The data of figure 4.7 and 4.8 were obtained by using the software Disperse [79].

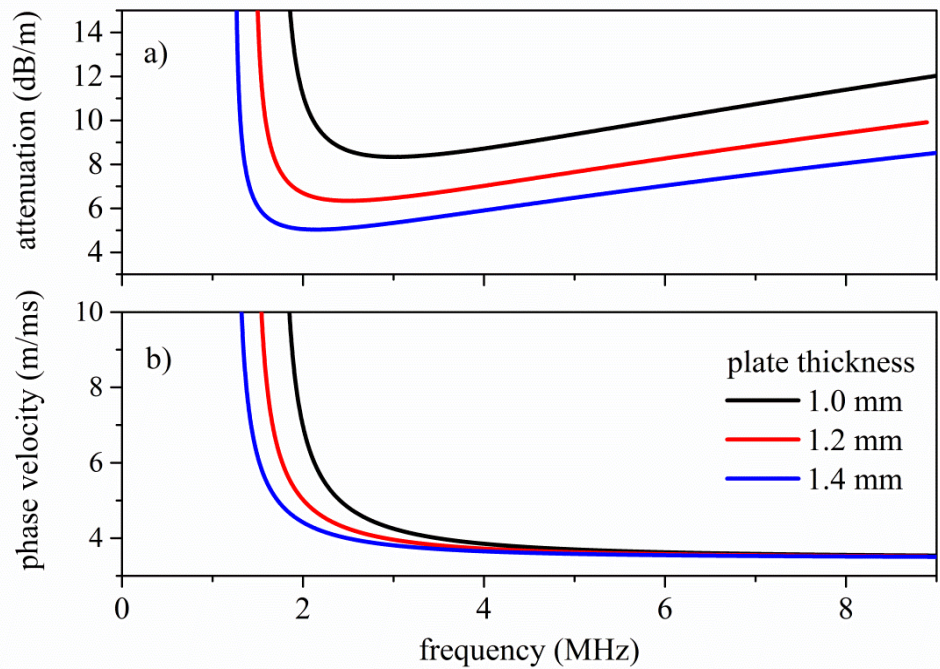

Figure 4.8: a) The attenuation and b) the phase velocity of the SH1 mode vs frequency for three different glass plate thicknesses: 1, 1.2 and 1.4 mm; the plate is immersed in kerosene.

Martin et al. [80] were the first to use the SHAPM device as a fluid phase sensor in 42.75° rotated Y-cut (RYC) quartz (ST-quartz): they experimentally verified the ability of the sensor to monitor the conditions at the solid/liquid interface. A bare quartz plate was used to measure the viscosity of water/glycerol mixtures, while the plate with the sensing surface chemically modified by ethylenediamine ligands was used to detect low concentrations of $Cu^{2+}$ ions in solution. After this paper, the SHAPM sensors have been successfully investigated for many applications. Some non-exhaustive examples of applications include the detection of mercury contamination in water, with (sub)-nanogram sensitivity, by using a ZX LNO and -65° Y rotated quartz plates covered by a gold sensitive membrane to accumulate the mercury via surface amalgamation [81]; the detection of potassium ions concentration in water with a relative frequency shift per unit potassium ions concentration was found equal to -8.37 ·$10^{-4}$ for the fundamental mode, by using a ST-90° x quartz plate covered with a polyvinyl-chloride-valinomycin membrane [77]; the detection of concentration of NaCl and tris(hydroxymethyl)aminomethane (Tris) in aqueous solution [82] or to analyse the surface density changes associated with cell adhesion and proliferation *in vitro* condition, by using a STx quartz plate [83].

In reference [78] experimental results with various SHAPMs in ST 90°-x quartz plate concerning the influence of the temperature, the viscosity and the concentration of NaCl and tris(hydroxymethyl)aminomethane (Tris) in aqueous solution are presented: the higher order modes appeared to be more sensitive than the first ones, although having more transmission losses.

## 5. Lamb Wave sensors

Lamb waves (LWs) are elastic guided waves that travel in finite thickness plates, between stress-free plane and parallel boundaries; they are elliptically polarized in that they show in-plane and out-of-plane particle displacement components, $U_1$ and $U_3$. We remand the reader to the book by Victorov [84] for LWs propagation details. LWs are divided into symmetric ($S_n$) and anti-symmetric ($A_n$) modes (where $n$ is the mode order). The former modes have the longitudinal displacement component $U_1$ that is symmetric with respect to the mid plane of the plate while $U_3$ is anti-symmetric; the opposite happens in the case of the anti-symmetric modes; figure 5.1 shows the total field profile and the single displacement components of the fundamental mode travelling along a Si plate.

.

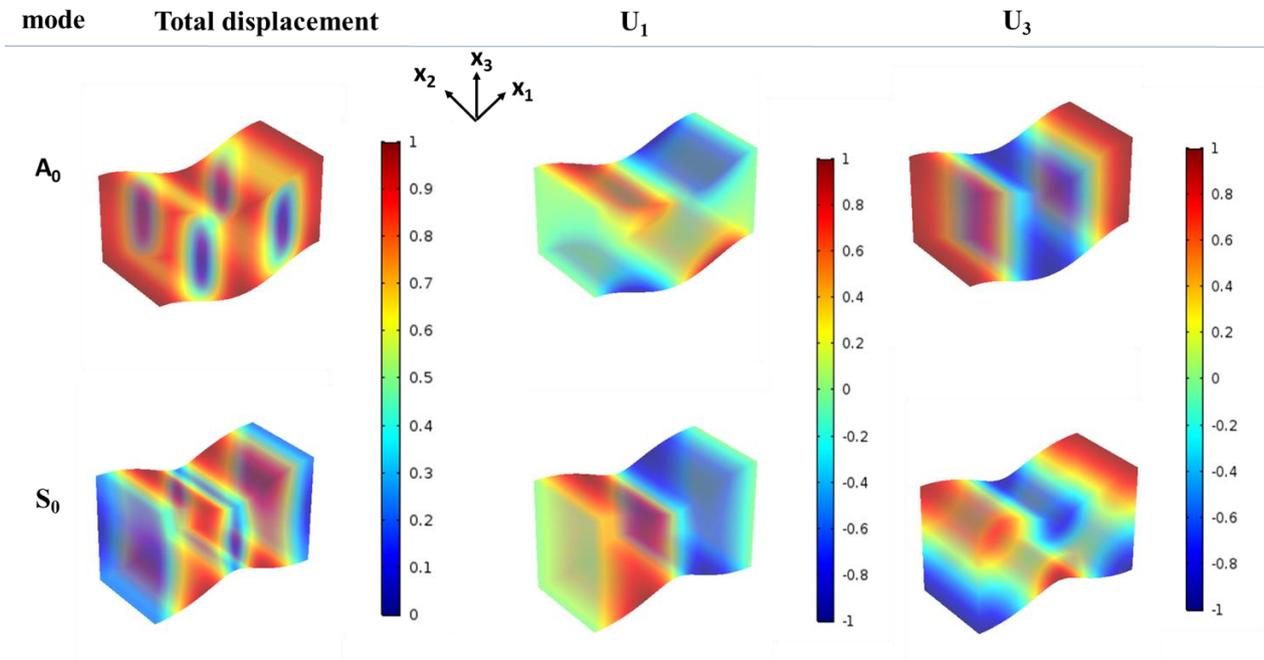

Figure 5.1 The total field profile and the two displacement components of the fundamental symmetric and anti-symmetric modes, $S_0$ and $A_0$ travelling along a Si plate.

The velocity of the modes depends on the plate characteristics (material type, thickness, crystallographic cut and wave propagation direction); the thicker the plate is, the more LWs modes exist. LWs are highly dispersive: as an example, figure 5.2 shows the phase velocity $v_{ph}$ vs the plate thickness-to-wavelength ratio curves of the symmetric $S_n$ (red curves) and anti-symmetric $A_n$ (blue curves) LWs travelling in a Si(001)<100> plate of thickness h. The shape of the modes, the displacement components variation across the cross section of the plate, changes considerably with the plate thickness and with the mode order [85]. The insets of figure 5.2 show the mode shape of the first six modes travelling along the plate with fixed thickness (h/$\lambda$ = 0.5); figure 5.3 shows the same $v_{ph}$ dispersion curves as in figure 5.2 but the insets are related to the shape of one mode ($S_2$) at different $h/\lambda$ (0.4. 1.0, and 1.8). The data of figures 5.2 and 5.3 were calculated using the DISPERSE software [79].

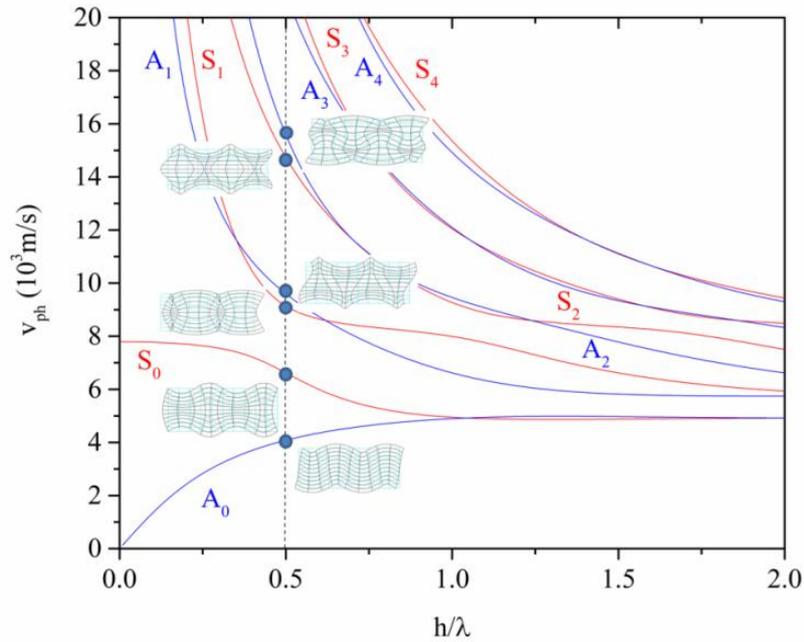

Figure 5.2: The $v_{ph}$ dispersion curves of the $S_n$ (red curves) and $A_n$ (blue curves) Lamb modes travelling in a Si(001)<100> plate in air. The insets show the field profile of different modes at the same abscissa value ($h/\lambda = 0.5$) and are marked by a blue dot.

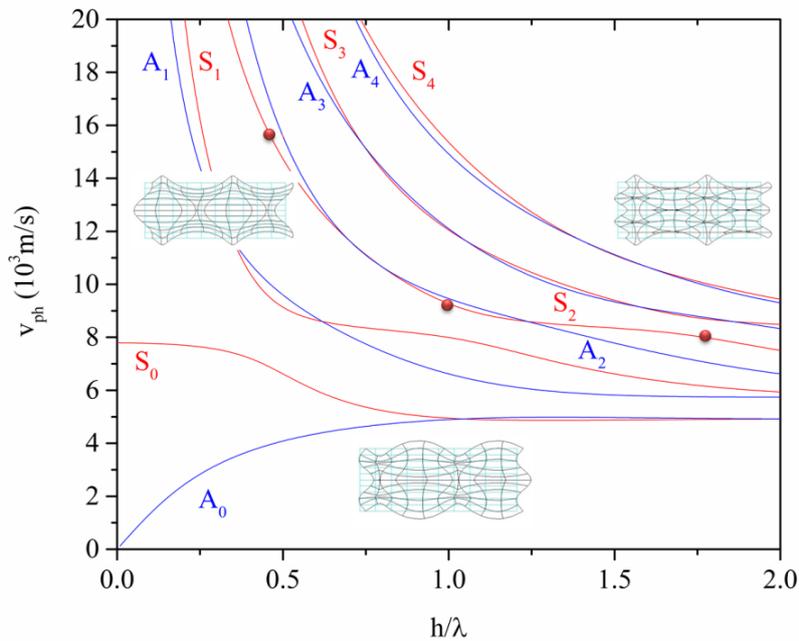

Figure 5.3: The $v_{ph}$ dispersion curves of the $S_n$ (red curves) and $A_n$ (blue curves) Lamb modes travelling in a Si(001)<100> plate in air. The insets show the field profile of the $S_2$ mode at different abscissa values (0.4. 1.0, and 1.8) marked by a red dot.

As the LWs have velocity higher than that of the surrounding liquid medium and have both in-plane and out-of-plane displacement components, they are not suitable for sensing applications in liquids, except in some special cases. These cases include: (1) a branch of the fundamental *symmetric* $S_0$ mode dispersion region where the longitudinal particle displacement component, $U_1$, is dominant over the out-of-plane component $U_3$ at both the plate surfaces and in the plate depth (the mode is mostly linearly polarized and propagates at a velocity slightly lower than the velocity of the longitudinal bulk acoustic wave, $v_{LBAW}$); (2) a branch of the higher order *symmetric* modes dispersion curve, where the modes have $U_3 \sim 0$ at the plate surfaces (but not in the plate depth), and travel at velocity equal to $v_{LBAW}$; (3) a branch of the fundamental *anti-symmetric* $A_0$ mode dispersion curve, to which corresponds a velocity lower than that of the fluid.

## 5.1 Quasi-longitudinal symmetric modes

Of great interest are certain points of the symmetric LWs dispersion curves where the phase velocity is close to the longitudinal bulk acoustic wave (LBAW) velocity of the plate material, $v_{LBAW}$, and the field profile has particular characteristics, such as $U_3 << U_1$, $U_2$ at the plate surfaces. These waves, named *quasi-longitudinal* LWs (QL-LWs), are able to travel along the surface of the plate while contacting a liquid environment without suffering large attenuation. Inside a small branch of the $S_0$ $v_{ph}$ dispersion curve, corresponding to $h/\lambda << 1$, $U_1$ can even have a constant amplitude along the whole depth of the plate, while $U_3$ is at least 10 times less than $U_1$ at any plate depth [86, 87]: the shape of the membrane particle movement is a flat ellipse and its longer axis is parallel to the surface of the plate. The higher order symmetric modes dispersion curves intersect the velocity of the LBAW in the plate material ($v_{LBAW} = 8440$ m/s for Si) and they show equal group velocity ($v_{gr} = 7275$ m/s). Figure 5.1.1 highlights the intersection of the LWs dispersion curves in a Si(001)<100> plate with the plate material $v_{LBAW}$: the mode shape of both the $A_n$ and $S_n$ modes at these points shows $U_3 = 0$ at the plate surfaces but, while the $A_n$ curves are highly dispersive, the $S_n$ modes show a flat dispersion region centred at the intersection point; this region corresponds to a $h/\lambda$ range where the condition $U_3 << U_1$ is satisfied, thus preventing the sensor performances to be highly affected by possible errors in the fabrication technology of the sensor device.

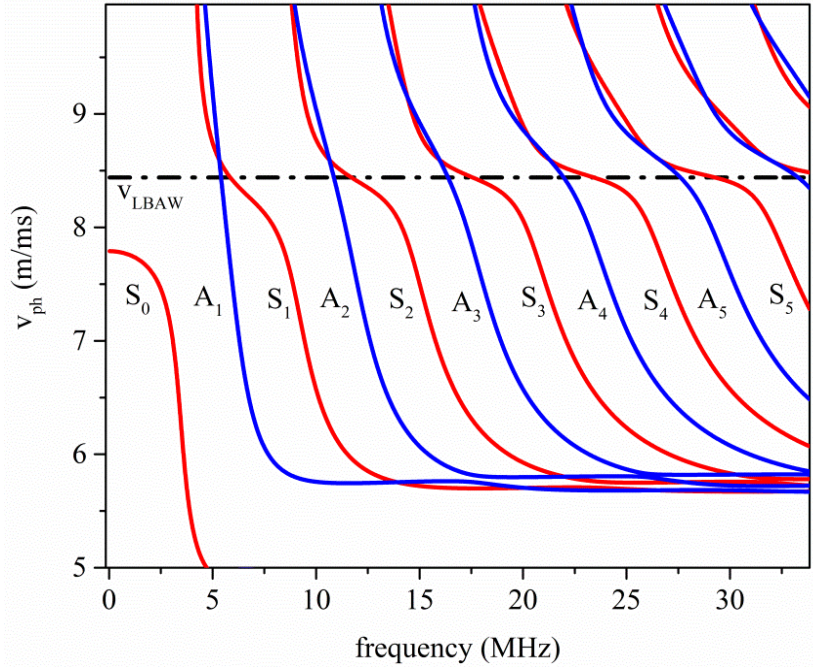

Figure 5.1.1: Dispersion curves for the LWs of a Si(001)<100> plate, 1mm thick, showing the intersections of the symmetric modes with a phase velocity equal to the LBAW velocity.

As an example figure 5.1.2 shows the field profile of the first four symmetric QL-LWs of figure 5.1.1: QL-$S_0$ (f = 0.441 MHz), QL-$S_1$ (f = 5.85 MHz), QL-$S_2$ (f = 11.70 MHz), and QL-$S_3$ (f = 17.56 MHz). For the QL-$S_1$ to QL-$S_3$ modes, $U_3$ is null only at the plate surfaces, but not inside the bulk of the plate; for QL-$S_0$ the $U_3$ vanishes on the plate surfaces and remains very small even in the plate depth, while $U_1$ is almost constant through the plate thickness.

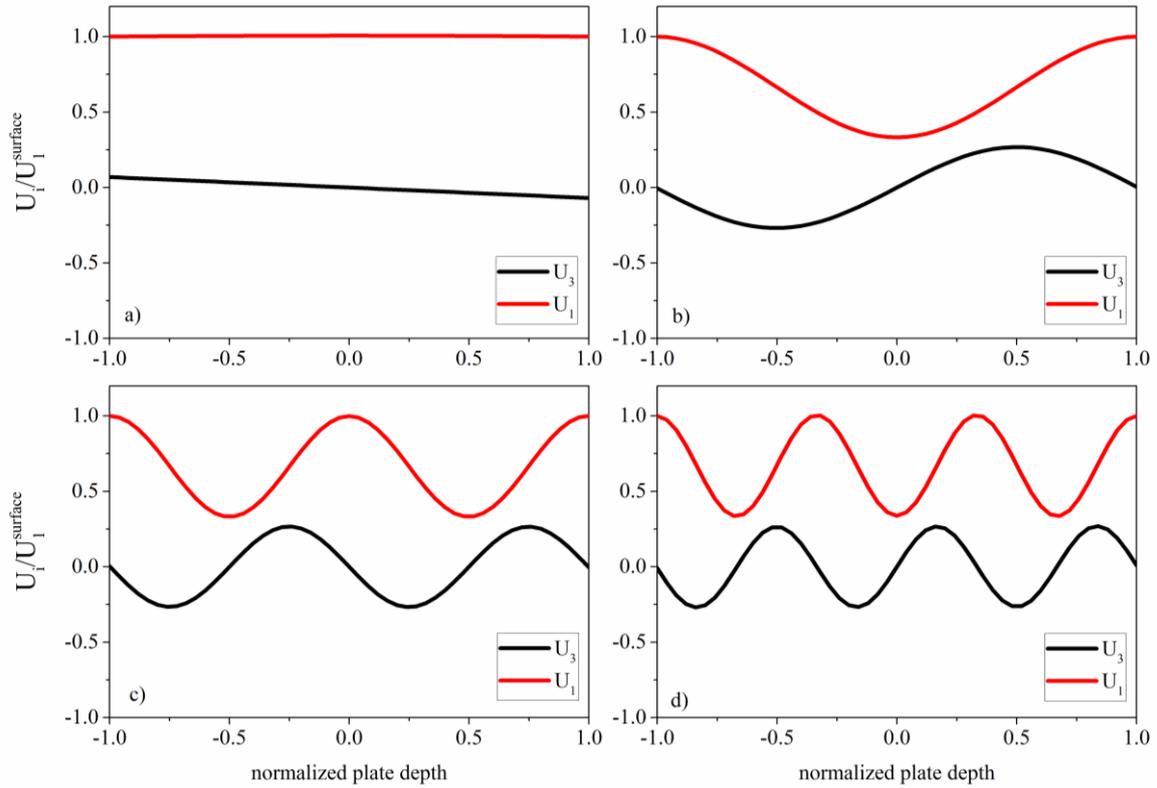

Figure 5.1.2: Cross-sectional normalized distribution of $U_1$ and $U_3$ displacement components in Si for: a) QL-S$_0$ at f = 0.441 MHz, b) QL-S$_1$ (f = 5.8219 MHz), c) QL-S$_2$ (f = 11.642 MHz), and d) QL-S$_3$ (f = 17.537 MHz).

As it can be seen in figure 5.1.2, the through-thickness for $U_1$ and $U_3$ are symmetric and antisymmetric about the mid plane of the plate, and the number of the minima increases with increasing the mode order. Since the $U_3$ component of these higher order modes vanishes on the free surfaces of the plate, these modes are suitable for liquid sensing.

Figure 5.1.3 shows the $v_{ph}$ and attenuation vs f·h curves for the first four symmetric modes (S$_0$, S$_1$, S$_2$ and S$_3$) for a Si plate immersed in water: when the velocity of the higher order modes reaches the $v_{LBAW}$ (8440 m/s), the attenuation rapidly drops to zero, thus confirming the modes suitability to sensing applications in liquid environments. The dispersion curves of figure 5.1.3 were normalized with respect to the plate thickness by plotting them against the frequency thickness product.

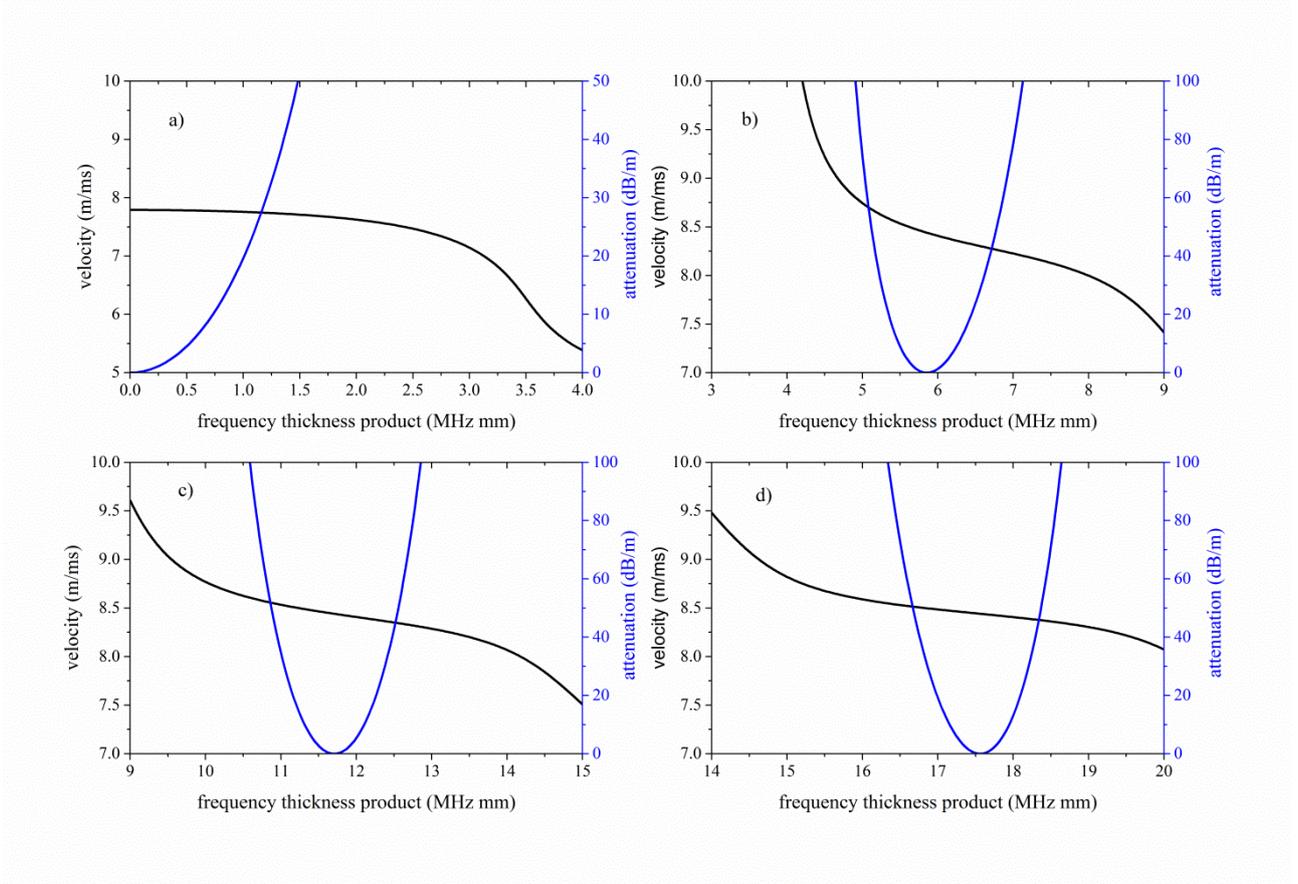

Figure 5.1.3: The $v_{ph}$ and attenuation vs f·h curves for the a) QL-$S_0$, b) QL-$S_1$, c) QL-$S_2$ and d) QL-$S_3$ modes travelling in a Si plate contacting a water half-space on both the two plate sides.

The fundamental mode QL-$S_0$ exhibits a plate normalized thickness value, $h/\lambda_{threshold}$, beyond which $U_3$ is no more negligible ($U_3 > 10\% \cdot U_1$ while $U_1 = 1$ at the plate surfaces but it is no longer constant inside the plate). For $h/\lambda < h/\lambda_{threshold}$, the wave has $U_1 = 1$ and constant along the plate depth, and the $U_3$ component is less than 10% of $U_1$.

The $h/\lambda_{threshold}$ has been calculated for several piezoelectric materials by using the McGill software [56] and the data are listed in table 5.1.1 [87]. For the higher order modes the $h/\lambda_{threshold}$ values listed in table 5.1.1 have a different meaning with respect to QL-$S_0$: it is the plate thickness corresponding to the minimum value of $U_3$ at the plate surfaces ($U_3/U_1 \sim 10^{-3}$), for $v_{ph} \sim v_{LBAW}$. By varying the thickness of the plate around $h/\lambda_{threshold}$, a $h/\lambda$ range ($h/\lambda_{range}$) can be found inside which the condition $U_3/U_1 \leq 0.1$ at the plate surfaces is verified. Table 5.1.1 summarizes the $h/\lambda_{threshold}$ and $h/\lambda_{range}$ of the fundamental and higher order quasi-longitudinal modes for some piezoelectric materials; the $K^2$ of the LWs have been evaluated for each material at the corresponding $h/\lambda_{threshold}$ for two coupling configurations.

Table 5.1.1: The $h/\lambda_{threshold}$, $h/\lambda_{range}$ and the $K^2$ values for two coupling configurations, for the QL-$S_0$ to QL-$S_3$ modes, for some piezoelectric materials. $h/\lambda_{range}$ represents the normalized thickness range, centered in $h/\lambda_{threshold}$ , where the condition $U_3/U_1 \leq 0.1$ at the plate surfaces is verified.

| material | | mode | | | |
|---|---|---|---|---|---|
| | | QL-$S_0$ | QL-$S_1$ | QL-$S_2$ | QL-$S_3$ |
| BN | h/$\lambda_{threshold}$ | 0.325 | 1 | 1.97 | 2.95 |
| | h/$\lambda_{range}$ | -- | 0.70-1.30 | 1.665 – 2.26 | 2.64 – 3.24 |
| | $K_{ST}^2$ ($K_{MST}^2$) (%) | 0.09 (0.14) | 0.035 (0.04) | 0.018 (0.020) | 0.012 (0.013) |
| ZnO | h/$\lambda_{threshold}$ | 0.07 | 0.65 | 1.24 | 1.86 |
| | h/$\lambda_{range}$ | -- | 0.59 – 0.68 | 1.17 – 1.305 | 1.81 - 1.925 |
| | $K_{ST}^2$ ($K_{MST}^2$) (%) | 0.47 (8.5) | 0.42 (0.50) | 0.22 (0.24) | 0.15 (0.16) |
| AlN | h/$\lambda_{threshold}$ | 0.11 | 0.79 | 1.58 | 2.37 |
| | h/$\lambda_{range}$ | -- | 0.75 - 0.89 | 1.5 – 1.67 | 2.29 – 2.46 |
| | $K_{ST}^2$ ($K_{MST}^2$) (%) | 0.35 (3) | 0.31 (0.37) | 0.17 (0.19) | 0.11 (0.13) |
| GaN | h/$\lambda_{threshold}$ | 0.12 | 0.77 | 1.53 | 2.29 |
| | h/$\lambda_{range}$ | -- | 0.735 – 0.86 | 1.51 – 1.62 | 2.2 – 2.38 |
| | $K_{ST}^2$ ($K_{MST}^2$) (%) | 0.26 (1.61) | 0.18 (0.20) | 0.095 (0.1) | 0.06 (0.07) |

The $K^2$ of both the two configurations are quite different for the QL-$S_0$ modes, but they become very similar with increasing the mode order. The $K^2$ decreases with increasing the mode order, and the MST configuration is always more efficient than the ST: this last effect is particularly evident for the QL-$S_0$ mode, while the $K_{ST}^2$ and $K_{MST}^2$ values become similar with increasing the mode order. Due to the small thickness value (h/$\lambda_{threshold}$ << 1) of the QL-$S_0$-based plates, the IDTs fingers are quite close to the opposite floating metal electrode for the MST configuration and consequently the electric field is mainly perpendicular to the plate surfaces. This results in a coupling efficiency quite larger than that of the ST configuration [87].

When Lamb waves travel along a non-homogenous plate (e.g. bi-layered composite plate), the symmetry of the particle displacement components with respect to the mid-plane of the plate is lost, unlike the homogeneous isotropic and anisotropic plates. The modes can be considered as quasi-$S_0$ and quasi-$A_0$ (q$S_0$ and q$A_0$) for a limited plate thickness range, while all the other modes can be generically labelled as i$^{th}$ mode. Unlike the single material plates, the $U_1$ and $U_3$ displacement components at the free surfaces of the composite plates can be quite different. As an example, figure 5.1.4a-c shows the field profile of three quasi longitudinal modes, q$S_0$, q$L_1$, and q$L_2$ travelling in AlN(1.4 μm)/SiN(0.2 μm) plate: the corresponding SiN/AlN total thicknesses values $H_{total}$/ $\lambda$ = ($h_{AlN}$ + $h_{SiN}$)/$\lambda$ values are 0.08, 0.80 and 1.6. As it can be seen, the condition $U_3$ << $U_1$ is verified on one plate side that is thus the one suitable for contacting a liquid environment [88].

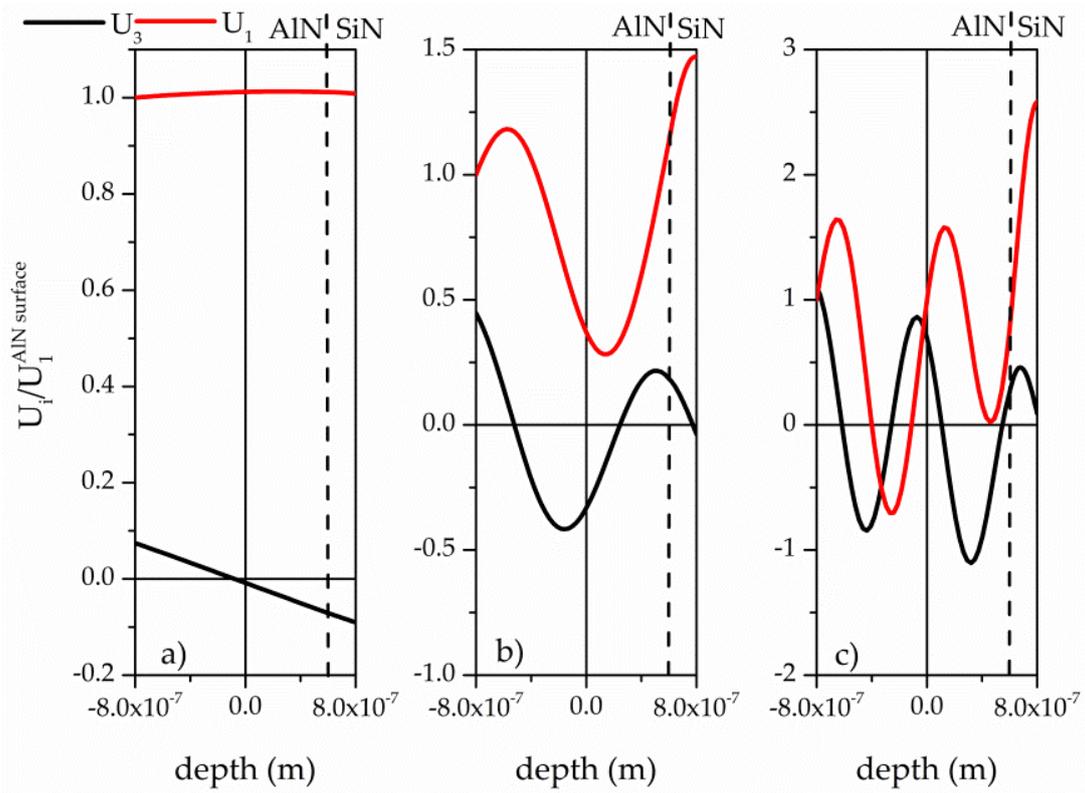

Figure 5.1.4: The field profile of the a) $qS_0$, b) $qL_1$, and c) $qL_2$ modes in air [88].

2D COMSOL Multiphysics software was employed to simulate the QL-$S_0$, QL-$S_1$ and QL-$S_2$ modes propagation along the AlN(1.4 μm)/SiN(0.2 μm) composite plate while contacting the liquid (water) environment from the SiN side of the plate where the $U_3 = 0$ is satisfied, as opposed to the AlN side of the plate. As it can be seen in figure 5.1.5, the acoustic energy remains confined inside the plate [88].

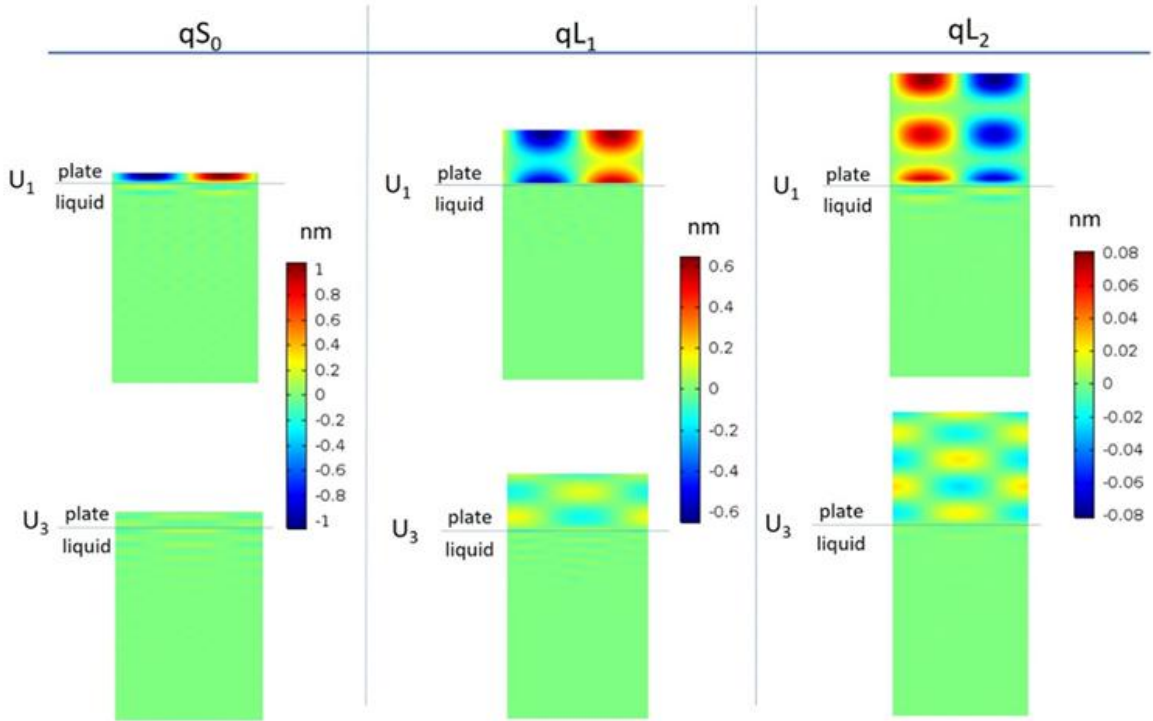

Figure 5.1.5: The FEM of the field profile for the $qS_0$, $qL_1$ and $qL_2$ modes at $(h_{AlN} + h_{SiN})/\lambda = 0.08$, 0.8 and 1.6, respectively [88].

In reference [89] the sensitivity to liquid viscosity of the QL-$S_0$ mode in wz-BN/c-AlN thin composite plates is theoretically predicted for different layers thicknesses. As an example, Table 5.1.2 lists the relative velocity shifts and the attenuation of the QL-$S_0$ mode in wBN/c-AlN composite plate contacting a liquid environment (70% of glycerol in water with $\rho_l = 1091.6$ kg/m$^3$, $\eta = 0.003$Pa·s, $\lambda = 10$ and 100 μm) for four different combinations of wBN and AlN thicknesses.

Table 5.1.2: The relative velocity shifts and the attenuation of the QL-$S_0$ mode in w-BN/c-AlN composite plate contacting a liquid environment (70% of glycerol in water with $\rho_l = 1091.6$ kg/m$^3$, $\eta = 0.003$Pa·s) for $\lambda = 10$ and 100 μm, for four different combinations of wBN and c-AlN thicknesses.

| QL-$S_0$ | | $\lambda = 10$ μm | | | $\lambda = 100$ μm | | |
|---|---|---|---|---|---|---|---|
| $h/\lambda_{wBN}$ | $h/\lambda_{c\text{-}AlN}$ | Frequency (GHz) | $\Delta v/v_0$ ($\cdot 10^{-4}$) | $\alpha$ (dB/cm) | Frequency (MHz) | $\Delta v/v_0$ ($\cdot 10^{-4}$) | $\alpha$ (dB/cm) |
| 0.1 | 0.013 | 1.6132 | -15.6 | -4.91 | 161.32 | -4.94621 | -0.155 |
| 0.2 | 0.01 | 1.647 | -8.70905 | -2.74 | 164.7 | -2.75404 | -0.086 |
| 0.3 | 0.004 | 1.6664 | -5.9056 | -1.85 | 166.64 | -1.86751 | -0.059 |
| 0.325 | 1E-3 | 1.871 | -4.95934 | -1.56 | 187.1 | -1.56828 | -0.049 |

## 5.2 Fundamental antisymmetric mode

Inside the LWs dispersion curves of figure 5.2, the $A_0$ mode is clearly identified by its reducing velocity as the plate thickness approaches zero. The $A_0$ mode, while being elliptically polarized, with $U_3$ not null at the plate surfaces, can travel along thin membranes that are in contact with a liquid if designed to travel at a velocity lower than that of most liquids, which lie in the range from 900 to about 1500 m/s, by choosing the proper plate thickness. At very small thickness-to-wavelength ratios, the phase velocity of the $A_0$ mode approaches zero; as the thickness increases, the velocity increases, and reaches asymptotically from below the SAW velocity of the plate material.

The Scholte mode, not shown in figure 5.1, is an anti-symmetric mode that propagates at the solid-fluid interface: its name comes from its similarity to the Scholte wave that is widely known in geophysics. The characteristic equation for the dispersion curve of this mode is obtained as a solution to the equations of continuity of stress and displacement at the solid-fluid interfaces to be solved for antisymmetric modes. In the low frequency limit, one solution is the $A_0$ mode and the other solution is the quasi-Scholte (Q-Sch): their velocity dispersion curves have a linear dependence with the frequency-thickness product. The dispersion curve of the latter mode is characterized by an asymptotic behaviour of the phase velocity approaching the sound speed in the fluid at high frequencies; this non-dispersive branch of the quasi-Scholte mode dispersion curve is named Scholte mode. The polarisation of the mode is mostly parallel to the interface with a small out-of-plane displacement component.

Figure 5.2.1 shows the phase velocity and attenuation curves versus the frequency for the $S_0$, $A_0$ and quasi-Scholte (Q-Sch) modes in a glass plate, 0.15 mm thick, immersed in water.

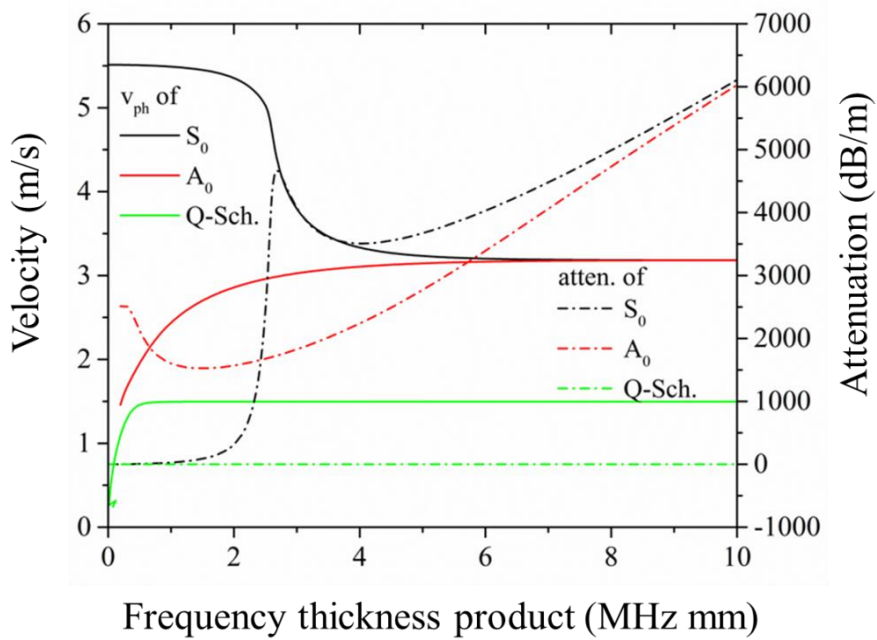

Figure 5.2.1: Phase velocity and attenuation dispersion curves of $S_0$, $A_0$ and Q-Sch modes for a glass plate 0.15 mm thick immersed in water.

The $S_0$ mode attenuation starts from zero with the frequency and reaches a local maximum value at 2.68 MHz·mm. In the limit as f·h tends to zero, the $S_0$ mode behaves basically as a longitudinally polarized wave, which explains its weak attenuation. When f·h tends to infinity the attenuation increases asymptotically with f $^2$: the mode is essentially concentrated at the surfaces of the plate and it radiates energy into the surrounding liquid in the same manner as a Rayleigh wave which is proportional to f². When the plate thickness becomes comparable with the acoustic wavelength, the $A_0$ mode behaves in the same way as the mode $S_0$ but it has an important attenuation in the low f ·h limit, caused by its flexural motion normal to the plate surface. The cut-off of the attenuation curve is at f·h =0.25 MHz·mm: below this limit, the phase velocity of the $A_0$ mode is smaller than the sound velocity in the liquid. As a result, no radiation of guided waves is allowed by the Snell law. The phase velocity of the Q-Sch plate mode rises with frequency from zero and gradually asymptotes to the velocity of the liquid half-spaces. Its attenuation is affected by the fluid bulk velocity, viscosity and bulk longitudinal attenuation. Q-Sch mode travels unattenuated in the direction of the phase velocity (if the fluid has no longitudinal attenuation); as it travels at velocity lower than the bulk velocity of the fluid, it is consequently evanescent in the direction orthogonal to the interface. The wave amplitude decays in an exponential manner with distance from the interface. The extent to which the wave penetrates into the fluid depends on the frequency, as shown in figure 5.2.2 where $U_1$, $U_3$ and the strain energy density (SED) of the Q-Sch mode travelling in a glass plate (1 mm thick) immersed in water at f = 107.602 and 454.186 kHz are shown.

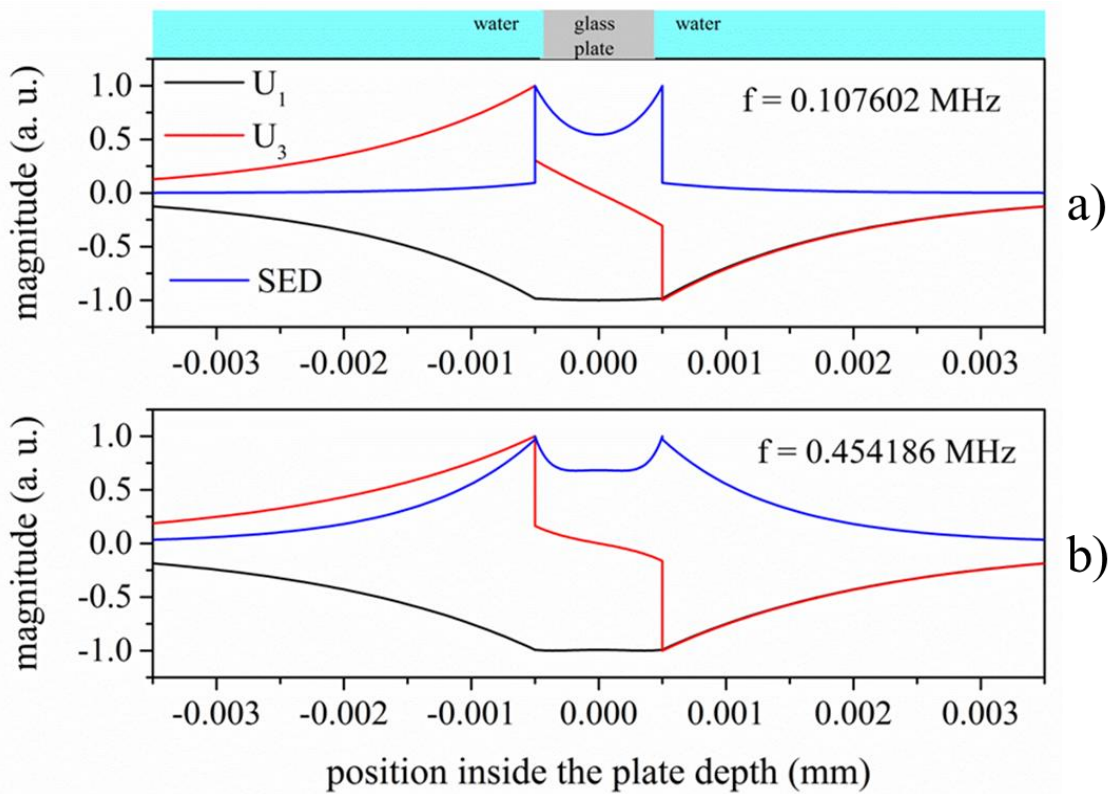

Figure 5.2.2: $U_1$, $U_3$ and the strain energy density (SED) of the Q-Sch mode travelling in a glass plate (1 mm thick) immersed in water at a) f = 107.602 and b) f = 454.186 kHz.

The Q-Sch mode energy distribution between the fluid and the plate depends on the frequency, as shown in figures 5.2.2 a and b: the out of plane displacement component at 107.602 kHz is almost constant across the section of the plate and the strain energy density indicates that the energy travels predominantly in the plate. At frequency 454.186 kHz a relevant part of the energy is travelling in the fluid. At higher frequencies (> 1MHz·mm) most of the energy travels in the fluid: the displacements decay away from the surfaces and are a minimum at the centre of the plate. Q-Sch waves can be used to characterize the fluid properties [90] since the wave attenuation, phase and group velocity are affected by the viscosity, longitudinal bulk attenuation and bulk velocity of the fluid. In reference [91] the influence of the waveguide material (steel, aluminium and brass) on the Q-Sch mode phase and group velocity sensitivities to the liquid parameters (longitudinal velocity and density) is theoretically studied. The study concluded that higher waveguide material density leads to higher sensitivities, and higher waveguide acoustic velocities lead to an extended effective sensing range.

As an example figure 5.2.3 shows Scholte mode phase and group velocity dispersion curves in Al plate (1 mm thick) immersed in water ($\rho$ = 1000 kg/m$^3$, v =1500 m/s), benzene ($\rho$ = 881 kg/m$^3$, v = 1117 m/s, dynamic viscosity $\eta$ = 0.65·10-3 Ns/m$^2$) and diesel ($\rho$ = 800 kg/m$^3$, v =1250 m/s ).

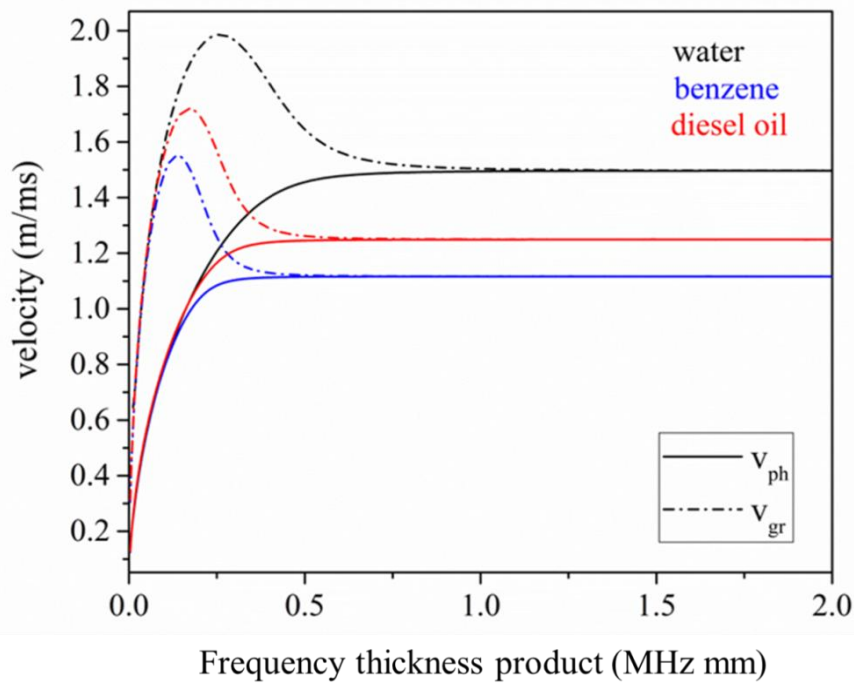

Figure 5.2.3: Scholte mode phase and group velocity dispersion curves in Al plate (1 mm thick) immersed in water ($\rho$ = 1000 kg/m$^3$, v =1500 m/s), benzene ($\rho$ = 881 kg/m$^3$, v = 1117 m/s, dynamic viscosity $\eta$ = 0.65·10-3 Ns/m$^2$) and diesel ($\rho$ = 800 kg/m$^3$, v =1250 m/s ).

In reference [92] the sensitivity of the quasi-Scholte mode for fluid characterization was assessed experimentally by measuring the phase velocity values for the quasi-Scholte mode in distilled water and in different ethanol-water concentrations. In reference [93] a preliminary sensitivity analysis is performed for application to simultaneous multi-sensing physical quantities of liquids such as temperature, viscosity and density using interface waves.

Figures 5.2.4a and b show the $A_0$ and $S_0$ mode velocity and attenuation dispersion curves in Si plate (1mm thick) immersed in glycerol ($\rho = 1258$ kg/m$^3$, v = 1860 m/s, dynamic viscosity $\eta = 1.49$ Ns/m$^2$), water ($\rho = 1000$ kg/m$^3$, v =1500 m/s), benzene ($\rho = 881$ kg/m$^3$, v = 1117 m/s, dynamic viscosity $\eta = 0.65 \cdot 10^{-3}$ Ns/m$^2$) and diesel ($\rho = 800$ kg/m$^3$, v =1250 m/s ); the longitudinal attenuation is set to zero.

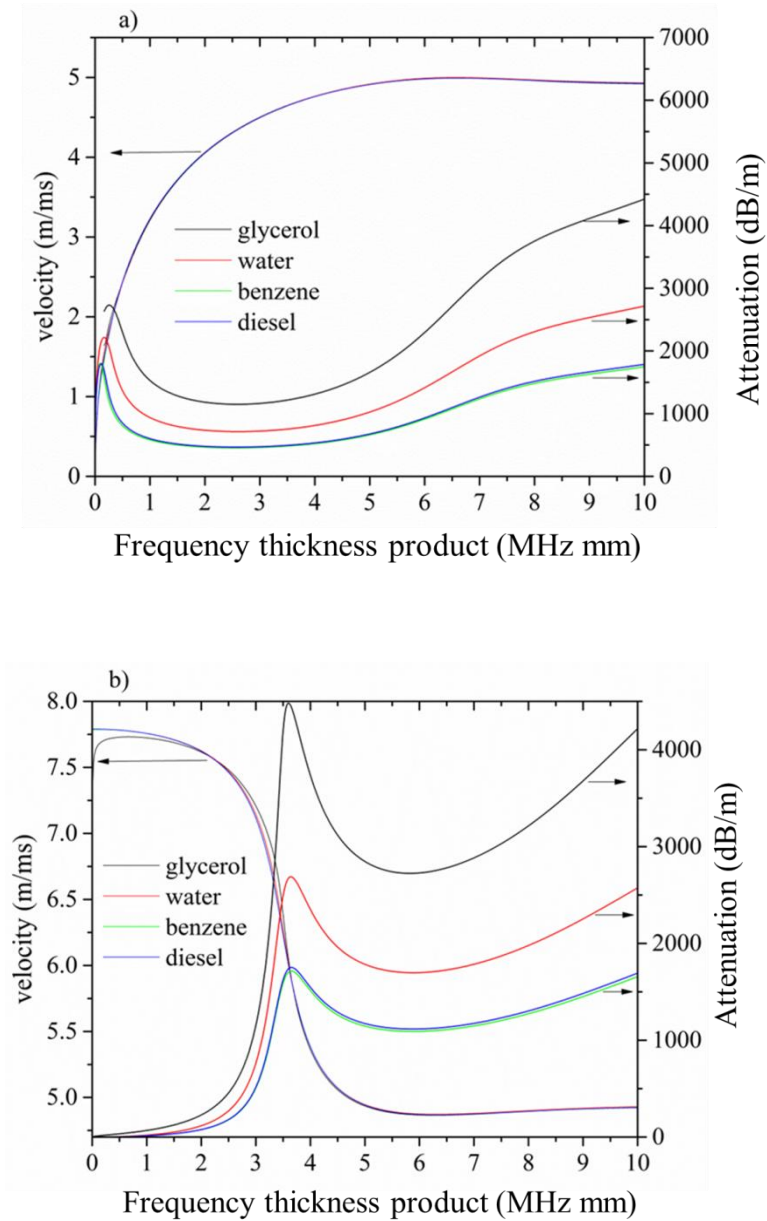

Figure 5.2.4: a) the $A_0$ mode and b) the $S_0$ mode velocity and attenuation dispersion curves in Si plate (1 mm thick) immersed in glycerol ($\rho = 1258$ kg/m$^3$, v = 1860 m/s, dynamic viscosity $\eta = 1.49$ Ns/m$^2$), water ($\rho = 1000$ kg/m$^3$, v =1500 m/s), benzene ($\rho = 881$ kg/m$^3$, v = 1117 m/s, dynamic viscosity $\eta = 0.65 \cdot 10^{-3}$ Ns/m$^2$) and diesel ($\rho = 800$ kg/m$^3$, v =1250 m/s ).

The $A_0$ and $S_0$ modes attenuation curves related to benzene and diesel are very close and difficult to distinguish; the same applies to the phase velocity curves but for all the studied liquids.

The $A_0$ mode attenuation is larger than that of the $S_0$ mode due to its faster decay over the propagation distance: the maximum attenuation (127.33 Nepers/m) happens when f·h equals 0.15 MHz mm, when the $A_0$ Lamb wave phase velocity is close to the water sound speed. When the $A_0$ phase velocity is less than the water sound speed, there is still attenuation that approaches to zero slowly as h/$\lambda$ comes to zero.

In the low viscosity range, the amplitude response of the sensor is also affected by other parameters, such as temperature, pressure and density which can play more important roles than the viscosity. In the case of water and diesel, the sensor responses to these two liquids are well distinguishable and are affected only by the mass density and velocity of the liquids, being the viscosity assumed to be equal to zero. On the contrary, the sensor responses to benzene and diesel, which have quite similar $\rho$ and $v_l$ but different (and very low) $\eta$, are very similar.

The devices based on the $A_0$ mode suffer some limitations, such as the low operating frequency (f = $v_{ph}/\lambda$) due to the $v_{ph}$ of $A_0$ mode which must be lower than the liquid velocity; thus the $A_0$-based device is not suitable to achieve high frequencies that is a prerequisite to enhance the sensor sensitivity [1]. If the device frequency is increased by reducing the IDT width, two technological problems are met: 1. if the sensor is implemented onto a single crystal piezoelectric substrate, the plate thickness must be scaled down together with the wavelength, thus increasing the fragility of the thinned plate; 2. If the sensor is implemented onto a thin suspended piezoelectric membrane, the layer structural quality imposes a limit to the maximum (and minimum) thickness. Another limitation is the achievable efficiency of electrical excitation of the acoustic wave. The $K^2$ of $A_0$ mode is dispersive as it depends on the membrane thickness. The theoretical $K^2$ dispersion curve of the $A_0$ mode is reported in reference [87] for various piezoelectric plates (BN, ZnO, InN, AlN and GaN), for two coupling structures: the Substrate/Transducer (ST) and Metal/Substrate/Transducer(MST) configurations. Table 5.2.1 lists the threshold plate thickness for operation in water ($v_{ph} \leq v_{water} = 1480$ m/s) and the corresponding $K^2$ of the ST and MST structures at the threshold [87].

Table 5.2.1: the $A_0$ mode h/$\lambda_{treshold}$ for operation in water ($v_{ph} \leq v_{water} = 1480$ m/s) and the corresponding $K^2$ of the ST and MST structures at h/$\lambda_{treshold}$.

| material | h/$\lambda_{treshold}$ | $K^2_{ST}$ (%) | $K^2_{MST}$ (%) |
|---|---|---|---|
| ZnO | 0.19 | 2.63 | 0.7 |
| BN | 0.05 | 0.011 | 0.0 |
| AlN | 0.087 | 0.25 | 0.02 |
| GaN | 0.12 | 0.35 | 0.06 |
| InN | 0.2 | 1.77 | 0.58 |

Even though the ZnO $K^2$ is relatively high, it is significantly smaller than that of the QL-$S_0$ mode (8.5% for MST structure).

The experimental result of fluid loading of a Lamb wave sensor employing $A_0$ mode was firstly reported by R.M White and S.W. Wenzel in 1988 [94]. Furthermore, the same group reported the experimental result of viscosity and density sensing using the same device [95] consisting of a composite SiN and ZnO membrane with the thicknesses range from 2.8 um to 6.0 µm and the IDT periodicity of 100 µm. In this configuration, the membrane normalized thickness is thin enough to obtain an $A_0$ phase velocity lower the sound velocity in water. For $h/\lambda = 0.06$, they obtained $A_0$ and $S_0$ modes with velocity of 470 m/s and 7850 m/s respectively. The effect of viscous fluid loading on $A_0$ mode is reported which show linear relationship between the attenuation loss and the square root of the product of fluid mass density and viscosity. Moreover, the authors demonstrated that simultaneous measurement of frequency shift and attenuation loss allows a fluid's viscosity and density to be determined.

In reference [96] experimental results on $A_0$-based sensor on PZT are described. The device was fabricated by deposition of low pressure chemical-vapor deposition (LPCVD) silicon nitride, 1 µm thick, a metal ground plane of Ta/Pt (10 nm/150 nm), and a 750-nm-thick layer of sol-gel-derived PZT on silicon wafer, followed by the lift off process of Ta/Pt (10 nm/150 nm) IDTs. The KOH was used for anisotropic etching of the back side of the silicon wafer to release the composite membranes structure. They obtained $A_0$ mode phase velocities in between 295–312 m/s and group velocities in between 414–454 m/s. A frequency shift of 850 kHz and an insertion loss as low as 3 dB are observed when the back side of the membrane is in contact with a column of 15-mm height of deionized water.

The theoretical study based on the Rayleigh's perturbation approach to compare the $A_0$ and $S_0$ Lamb wave sensors sensitivity in liquid is reported in reference [97]: it is shown that the sensitivity of the $A_0$ mode is much greater than that of $S_0$ mode. In reference [98] the experimental test of the $A_0$, $S_0$ and $SH_0$ sensors demonstrates that the $A_0$ mode frequency shift caused by the presence of liquid is quite larger than that of the $S_0$ and $SH_0$ modes.

The most recent experimental result of $S_0$ mode to measure the mechanical and electrical liquid properties is reported by Miera et. Al. [99]: using two AlN-based $S_0$ sensor topologies, with floating bottom metallic layer and without, the influence of mechanical and electrical properties of different aqueous mixtures was experimentally assessed.

## 6. Discussion and conclusions

All acoustic wave devices behave like sensors since they are highly sensitive to any change in the boundary conditions that result in a wave velocity and/or propagation loss change. If the acoustic wave path is covered by a sensitive coating that is able to absorb only specific chemical vapors or specific biological chemicals in liquids, the sensor becomes a chemical sensor or a biosensor. All the acoustic wave sensors are able to work in gaseous environments, but only a subset of them can operate in contact with liquids. The waves that are predominantly in-plane polarized do not radiate appreciable energy into liquids contacting the device surface, as opposed to those waves with a substantial out-of-plane displacement component, which radiates compressive waves into the liquid, thus causing excessive damping. An exception to this rule occurs for devices utilizing waves that propagate at a velocity lower than the sound velocity in the liquid.

The quartz crystal microbalance (QCM) is one of the most common shear horizontal mode sensor: it typically consists of a thin disk of AT-cut quartz with parallel metal electrodes patterned on both sides. When a voltage is applied between these electrodes, the plate undergoes a shear deformation. The QCM resonant frequency, $f = v_{BAW}/2h$ , is inversely proportional to the quartz plate thickness h (few hundreds of micrometers): its value is typically between 5–30 MHz. When the QCM contacts a Newtonian liquid, it results in a resonance frequency shift proportional to the square root of the liquid viscosity-density product, $\Delta f = -f_0^{3/2} \sqrt{\rho_l \eta_l / \rho_q \mu_q \pi}$ , as described by Kanazawa and Gordon [100]; $\rho_L$ and $\eta_L$ denote the density and viscosity of the liquid, $\rho_Q$ is the density of quartz, $\mu_Q$ is the elastic constant of the piezoelectrically stiffened quartz, $f_0$ is the QCM resonant frequency, respectively. The fluid is entrained within a distance from the quartz surface that is equal to the wave penetration depth, $\delta = \sqrt{2\eta_l / \omega \rho_l}$. An increased QCM sensitivity requires a higher resonant frequency: this effect can be achieved by thinning the quartz plate that thus becomes more fragile and difficult to manufacture and handle.

An alternative to the QCM are the *surface-generated* acoustic wave sensors whose operating frequency is determined by the periodicity of the IDTs and the mode velocity, $f = v/\lambda$. Moreover their acoustic energy is trapped near the surface where the sensing phenomena take place: the more acoustic energy is concentrated at the surface, the higher the sensitivity to surface perturbations. PSAW, HVPSAW, Love waves, and Lamb waves belong to this group. SHAPMs modes do not belong to this group since the acoustic energy is distributed throughout the bulk of the substrate, even if the waves propagation is excited and detected by means of the IDTs, as well as for the SAWs. The sensitivity of the SHAPM sensors depends strongly on the thickness of the substrate: a higher sensitivity can be obtained by thinning the substrate, at the cost of lack of substrate robustness. These sensors are more sensitive than the QCM, but less sensitive than the other surface-generated acoustic wave sensors since the energy of the wave is distributed throughout the bulk of the substrate and not trapped at the surface where the sensing phenomena take place. An advantageous feature of the SHAPM sensors is that these devices are sensitive on both sides of the substrate so that the back side can be used for sensing while having the front protected from the liquid [86]. This simple technology is opposed to that required by the Lamb Wave sensors, that requires the thinning of the bulk piezoelectric substrate or the release of a thin suspended membrane, or to that of the Love wave sensors, that require the growing of a guiding layer on top of the piezoelectric half-space.

Thin piezoelectric suspended membranes are fabricated by employing the silicon micromachining techniques. By using the thin piezoelectric film technology, high-frequency devices can be designed that offer an important advantage: the surrounding electronic circuitry (amplifier) can be integrated with the acoustic device, thus offering the possibility for on-chip compensation of disturbing effects such as temperature or pressure variation. The fundamental anti-symmetric Lamb mode travelling along a very thin membrane (thickness typically 5% or less of the wavelength) has a phase velocity lower than the sound velocity of the loading liquid $v_l$ (typically from 900 to ~1900 m/s). Therefore, the plate functions as a wave guide yielding no radiation losses despite the wave has a non-null shear vertical particle displacement component. Due to the low phase velocity of the $A_0$ mode, the operating frequency is typically in the range of 5-20 MHz. The fundamental symmetric Lamb mode, $S_0$, is predominantly longitudinally polarized just for small plate thickness values: as its phase velocity is much higher than that of the $A_0$ mode, it can reach higher sensibility.

The higher order quasi-longitudinal Lamb modes are promising candidates for liquid sensing applications: they have high velocity (close to that of the longitudinal BAW in the plate material) and require a fabrication technology simpler than that of the other Lamb wave sensors since they correspond a thicker plate thickness-to-wavelength ratio. For example, $h/\lambda = 0.07$ is the upper limit of the ZnO plate thickness that allows the propagation of longitudinally polarized $S_0$ mode, with $K^2$ = 9%. Higher order quasi-symmetric modes $S_1$, $S_2$ $S_3$, and $S_4$, propagate for $h/\lambda \sim 0.65$, 1.24, 1.86, 2.48. Their $K^2$ is equal to 0.42%, 0.22%, 0.15%, and 0.10%, respectively, much lower than that of the $S_0$ mode. For 10 μm thick ZnO membrane, the $S_1$, $S_2$, $S_3$ and $S_4$-based devices operating frequencies are 394 MHz, 754 MHz and 1125 MHz and 1507 MHz, respectively, as opposed to that (38 MHz) estimated for the $S_0$ mode [101]. The thin suspended membrane devices may be only a few micrometers thick, thus the mass per unit area of the thin plate can be increased significantly by the mass-loading effect produced by the changes in the density of a fluid, or by the attachment of protein molecules, cells, and bacteria from a liquid onto the suspended membrane surface. Moreover, these sensors can be fabricated by techniques compatible with planar integrated circuit technology thus allowing the low cost and small-sized sensors fabrication and the integration of the device with the surrounding electronic circuitry.

Love wave devices include a wave-guiding film deposited on top of a substrate [102]. The sensitivity of the Love wave sensors depends on the layer and half-space materials type and their crystallographic orientation, and on the layer thickness. There is a relationship between the slope of the dispersion curve to the mass sensitivity of a Love wave sensor that allows to find the layer thickness corresponding to the optimal sensor sensitivity. If the Love wave sensor consists in a piezoelectric layer covering a non-piezoelectric half-space, then the IDTs can be buried under the guiding layer that recovers the additional role of shielding the IDTs from aggressive liquid environment. The Love wave sensor can also include a sensing coating deposited onto the waveguiding layer: if this coating layer has the proper elastic properties, then it can also cover the role of waveguiding layer, with the advantage of avoiding a step of the device manufacturing process.

If the cut of the piezoelectric crystal substrate is properly rotated, the wave propagation mode changes from an elliptically polarized SAW to in-plane polarized SAWs, that are the PSAW and HVPSAW. This dramatically reduces the losses when liquids contact the propagating medium, thus allowing these devices to operate as biosensors. The PSAW and HVPSAW phase velocities are higher than the generalized SAW phase velocity, thus allowing the fabrication of higher frequency devices for the same photolithographic resolution, which also increases the sensor sensitivity.The PSAW and HVPSAW-based sensors have many advantages, such as a simple structure, highly reproducible large-scale fabrication technology, straightforward integration with microfluidic channels; moreover, their operating frequencies are higher than the QCM counterpart. However, a disadvantage of such devices is that the IDTs are located on the same side of the substrate that contacts the liquid: as a consequence, the test cell must have a reduced size and must be sealed within the propagation path.

Provided that the LWs, SHAPMs, the Love modes and the PSAW-based devices are good candidates for biosensing and chemical sensing in liquids, it is worth noting to underline that the design parameters (such as materials type, crystallographic orientation, electrical boundary conditions, layers thickness, etc.) of any electroacoustic device highly affect the sensors performances. It has been proven that numerical calculations and FEM analysis yield an in-depth study of the SAW characteristics to obtain very useful description of the sensor's performances and a deep assessment of the device sensitivity.

The present study wants to point the reader's attention towards the features of different kinds of acoustic waves and modes, as well as optimal combinations of materials and electrode structures, that can be exploited to design electro-acoustic devices suitable for chemical and biochemical sensing applications. A promising application of the SAW-based devices consist of a fully integrated platform that uses the SAWs to develop several processes, from sample handling to detection and measure. SAW-directed transport of analytes can be coupled to the SAW sensing functionalities. The SAW microfluidic systems can move and remove liquid targets to and from different sensing modules on the same chip; SAW sensors can make analysis of fluids as well as separation and fractionation of cells. However, it is important to point out that the future developments in acoustic-type chemical and biological sensors is also strongly related to the development of sensitive membranes which are responsible for the selectivity of the sensors: at this aim, coordinated efforts between researchers in diverse disciplines (such as electrical engineering, physics, chemistry, microbiology and medicine) are needed.

## Acknowledgment

This project has received funding from the European Union's Horizon 2020 research and innovation programme under the Marie Sklodowska-Curie Grant Agreement No. 642688.

## ORCID iDs

C Caliendo https://orcid.org/0000-0002-8363-2972

M Hamidullah https://orcid.org/0000-0003-2131-4335